\documentclass{amsart}
\usepackage{latexsym}
\usepackage{amsmath}
\usepackage{amsfonts}
\usepackage{amssymb}
\usepackage{graphicx}
\newtheorem{theorem}{Theorem}

\newtheorem{definition}[theorem]{Definition}
\newtheorem{example}[theorem]{Example}

\newtheorem{problem}{Open Problem}
\newtheorem{proposition}[theorem]{Proposition}

\let\frak=\mathfrak

\begin{document}

\author{Brian C. Hall}

\address{Department of Mathematics\\University of Notre Dame\\Notre Dame, IN 46556 U.S.A.}

\email{bhall@nd.edu}

\renewcommand{\subjclassname}{\textup{2000} Mathematics Subject Classification}

\title{Harmonic analysis with respect to heat kernel measure}
\date{June, 2000}
\begin{abstract}
I survey developments over the last decade in harmonic analysis on Lie groups
relative to a heat kernel measure. These include analogs of the Hermite
expansion, the Segal-Bargmann transform, and the Taylor expansion. Some of the
results can be understood from the standpoint of geometric quantization.
Others are intimately related to stochastic analysis.
\end{abstract}
\maketitle
\tableofcontents

\section{Introduction}

Parts of harmonic analysis on Euclidean spaces can naturally be
expressed in terms of a Gaussian measure, that is, a measure of the form
$a\exp(-b\left|  x\right|  ^{2})\,dx,$ where $dx$ is Lebesgue measure and $a$
and $b$ are positive constants. Among these are logarithmic Sobolev
inequalities, Hermite expansions, the quantum harmonic oscillator, coherent
states, and the Segal--Bargmann transform. Expressing things in terms of
Gaussian measure rather than Lebesgue measure also permits one to let the
dimension tend to infinity, allowing for example, an infinite-dimensional
version of the Fourier transform (the Fourier--Wiener transform) and of the
Hermite expansion (the chaos expansion, a basic tool in stochastic analysis).

There have been various constructions over the years that allow similar
developments on manifolds other than Euclidean space. This article summarizes
one set of such constructions, in which Euclidean space is replaced by a Lie
group and the Gaussian measure on Euclidean space is replaced by a
\textit{heat kernel measure} on the Lie group. The results that I will
describe here had its beginning with work of L. Gross in 1993, and the theory
has been developed over the last decade by Gross and many other authors. I
will describe a number of these results as well as several open problems.

The original results of Gross were intimately related to stochastic analysis,
specifically, analysis on loop groups. Although most of the results described
in this survey can be understood purely non-probabilistically, several other
connections with stochastic analysis have emerged. (See Section 4.) In another
direction, results involving heat kernels on groups have provided surprising
new examples in geometric quantization, including the phenomenon of
``quantization commuting with reduction.'' (See Section 3 and Section 4.2.)
The methods described here have also been used to give new insights into the
quantization of 2-dimensional Yang-Mills theory. (See Section 4.2.) Many of
the results here involve natural ``creation and annihilation operators''; a
challenging open problem is to understand these results in terms of some
variant of the canonical commutation relations and the Stone-von Neumann
theorem. (See Section 2.5.) It is my hope that this survey will draw attention
to some of the interesting open problems in this area.

In this introduction I will summarize some of the results for the Euclidean
case. I will concentrate on three different Hilbert spaces and the natural
unitary isomorphisms among them. For a more detailed exposition of some of
these results see \cite{H6,F}. Although the terminology comes from the
mathematical physics literature, these constructions are also an important
part of harmonic analysis. All of the constructions depend on a positive
parameter $t,$ which may be interpreted as Planck's constant.

Three Hilbert spaces are involved. The first is the \textbf{position Hilbert
space}, $L^{2}(\mathbb{R}^{d},\rho_{t}).$ Here $\rho_{t}$ is the normalized
Gaussian measure given by
\begin{equation}
d\rho_{t}\left(  x\right)  =\left(  2\pi t\right)  ^{-d/2}e^{-x^{2}%
/2t}\,dx.\label{gauss.def1}%
\end{equation}
Here $x^{2}=x_{1}^{2}+\cdots+x_{d}^{2}.$ The use of Gaussian measure instead
of Lebesgue measure is merely a convenience that simplifies many of the
constructions described here. As long as the dimension $d$ remains finite, one
can always translate all results back to the Lebesgue measure setting.

The second Hilbert space is the \textbf{Segal-Bargmann space} $\mathcal{H}%
L^{2}(\mathbb{C}^{d},\mu_{t}).$ Here $\mathcal{H}L^{2}(\mathbb{C}^{d},\mu
_{t})$ denotes the space of entire holomorphic functions on $\mathbb{C}^{d}$
that are square-integrable with respect to the measure $\mu_{t}$ given by
\begin{equation}
d\mu_{t}\left(  z\right)  =\left(  \pi t\right)  ^{-d/2}e^{-\left|  z\right|
^{2}/t}\,dz.\label{gauss.def2}%
\end{equation}
Here $dz$ denotes $2d$-dimensional Lebesgue measure on $\mathbb{C}^{d}$.

The third Hilbert space is what I will call the \textbf{Fock space.} (The
Segal-Bargmann space is also sometimes called ``Fock space.'') I will give
here a slightly unorthodox description of the Fock space, which will be
convenient later. Let $S(\mathbb{R}^{d})$ denote the symmetric algebra over
$\mathbb{R}^{d}.$ The symmetric algebra is just the tensor algebra
$T(\mathbb{R}^{d}),$ modulo the two-sided ideal $I$ generated by elements of
the form $X\otimes Y-Y\otimes X,$ with $X,Y\in\mathbb{R}^{d}.$ Now let
$I^{0}(\mathbb{R}^{d})$ denote the space of complex-valued linear functionals
on $T(\mathbb{R}^{d})$ that are zero on the ideal $I.$ So $I^{0}%
(\mathbb{R}^{d})$ is just the complex-valued dual space to the symmetric
algebra $S(\mathbb{R}^{d})=T(\mathbb{R}^{d})/I.$ A typical element $\xi$ of
$I^{0}(\mathbb{R}^{d})$ is a sum
\begin{equation}
\xi=\xi_{0}+\xi_{1}+\xi_{2}+\cdots,\label{xi.form}%
\end{equation}
where each $\xi_{n}$ is a symmetric element of the dual of $(\mathbb{R}%
^{d})^{\otimes n}.$ We make no restriction at the moment on the $\xi_{n}$'s;
in particular infinitely many of them can be non-zero.

Now let $X_{1},\cdots,X_{d}$ be an orthonormal basis for $\mathbb{R}^{d}$ with
respect to the standard inner product. Define a norm $\left\|  \cdot\right\|
_{n}$ on the dual of $(\mathbb{R}^{d})^{\otimes n}$ by setting
\begin{equation}
\left\|  \xi_{n}\right\|  _{n}^{2}=\sum_{k_{1},\cdots,k_{n}=1}^{d}\left|
\xi_{n}\left(  X_{k_{1}}\otimes\cdots\otimes X_{k_{n}}\right)  \right|
^{2}.\label{xin.norm}%
\end{equation}
This norm is independent of the choice of orthonormal basis for $\mathbb{R}%
^{d}. $ Next define a norm $\left\|  \cdot\right\|  _{t}$ on $I^{0}%
(\mathbb{R}^{d})$ by setting
\begin{equation}
\left\|  \xi\right\|  _{t}^{2}=\sum_{n=0}^{\infty}\frac{t^{n}}{n!}\left\|
\xi_{n}\right\|  _{n}^{2},\label{fock.norm}%
\end{equation}
where the $\xi_{n}$'s are as in (\ref{xi.form}). Then the Fock space
$I_{t}^{0}(\mathbb{R}^{d})$ is defined as
\[
I_{t}^{0}(\mathbb{R}^{d})=\left\{  \left.  \xi\in I^{0}(\mathbb{R}%
^{d})\right|  \left\|  \xi\right\|  _{t}<\infty\right\}  .
\]

There are natural unitary maps between each pair of Hilbert spaces described
above. These unitary maps allow one to transfer problems from one space to
another; each of the three spaces has advantages for certain types of
problems. The position Hilbert space is how we are most accustomed to thinking
about things. The Segal-Bargmann space is a sort of ``phase space'' Hilbert
space that incorporates simultaneously information about a function and its
Fourier transform, or in quantum-mechanical lingo, information about the
position and the momentum of a particle. The Fock space encodes the
decomposition of a function into ``excited states'' for a quantum harmonic
oscillator. In the infinite-dimensional theory ($d\rightarrow\infty$), the
Fock space encodes the decomposition into states describing a fixed number of
particles. Besides the position Hilbert space, Segal-Bargmann space, and Fock
space there is also the momentum Hilbert space, which is obtained from the
position Hilbert space by means of the Fourier transform. I will not discuss
further the momentum Hilbert space in this survey.

These three Hilbert spaces, and the isomorphisms between them, can be
understood in terms of the ``creation and annihilation operators'' and the
``canonical commutation relations.'' This point of view is discussed in
Section 2.5 and, among other places, in \cite{H6}.

First, between the position Hilbert space and the Segal-Bargmann space we have
the \textbf{Segal-Bargmann transform}. For $f\in L^{2}(\mathbb{R}^{d},\rho
_{t})$ we define the Segal-Bargmann transform $B_{t}f$ of $f$ by
\begin{equation}
B_{t}f\left(  z\right)  =(2\pi t)^{-d/2}\int_{\mathbb{R}^{d}}e^{-(z-x)^{2}%
/2t}f(x)\,dx,\quad z\in\mathbb{C}^{d}.\label{bt.form1}%
\end{equation}
Here $\left(  z-x\right)  ^{2}=\left(  z_{1}-x_{1}\right)  ^{2}+\cdots+\left(
z_{d}-x_{d}\right)  ^{2}.$ It is not hard to see that for each $f\in
L^{2}(\mathbb{R}^{d},\rho_{t}),$ the integral in (\ref{bt.form1}) converges
absolutely and the result is a holomorphic function of $z\in\mathbb{C}^{d}.$
Note that if we restrict attention to $z\in\mathbb{R}^{d},$ then
$B_{t}f\left(  z\right)  $ is just the convolution of $f$ with a Gaussian.
This is just the solution of the heat equation at time $t$ with initial
condition $f.$ Thus we may write
\begin{equation}
B_{t}f=\text{ analytic continuation of }e^{t\Delta/2}f.\label{bt.form2}%
\end{equation}
Here $e^{t\Delta/2}f$ is a mnemonic for the solution of the heat equation with
initial function $f,$ and the analytic continuation is from $\mathbb{R}^{d} $
to $\mathbb{C}^{d}$ with $t$ fixed. My convention is that the Laplacian is a
negative operator, so that $e^{t\Delta/2}$ is the \textit{forward} heat operator.

A theorem of Segal and Bargmann is the following.

\begin{theorem}
[Segal-Bargmann]For all $t>0,$ the map $B_{t}$ is a unitary map of
$L^{2}(\mathbb{R}^{d},\rho_{t})$ onto $\mathcal{H}L^{2}(\mathbb{C}^{d},\mu_{t}).$
\end{theorem}

\noindent See \cite{B,H6,BSZ,F} for more information.

Next we consider the \textbf{Taylor map} between the Segal-Bargmann space
$\mathcal{H}L^{2}(\mathbb{C}^{d},\mu_{t})$ and the Fock space. Let us consider
a more concrete description of the Fock space. Suppose $\xi=\xi_{0}+\xi
_{1}+\cdots$ is an element of the Fock space $I_{t}^{0},$ as in (\ref{xi.form}%
). Then each $\xi_{n}$ is a complex-valued linear functional on $(\mathbb{R}%
^{d})^{\otimes n}.$ So if $X_{1},\cdots,X_{d}$ is an orthonormal basis for
$\mathbb{R}^{d}$ then $\xi_{n}$ is determined by the complex numbers
\[
\alpha_{k_{1},\cdots,k_{n}}:=\xi_{n}\left(  X_{k_{1}}\otimes\cdots\otimes
X_{k_{n}}\right)  .
\]
Since $\xi$ annihilates the ideal $I,$ each $\xi_{n}$ must be a symmetric
linear functional, so the numbers $\alpha_{k_{1},\cdots,k_{n}}$ must be
invariant under permutation of the indices $k_{1},\cdots,k_{n}.$ The norm of
$\xi$ (as given in (\ref{fock.norm})) may be expressed in terms of the
$\alpha_{k_{1},\cdots,k_{n}}$'s as
\begin{equation}
\left\|  \xi\right\|  _{t}^{2}=\sum_{n=0}^{\infty}\frac{t^{n}}{n!}\sum
_{k_{1},\cdots,k_{n}}\left|  \alpha_{k_{1},\cdots,k_{n}}\right|
^{2}.\label{fock.norm1}%
\end{equation}
Since the $\alpha$'s are symmetric this may be rewritten in a way that
involves only sets of indices with $k_{1}\leq\cdots\leq k_{n},$ with some
additional combinatorial factors. However, for purposes of generalization to
the group case, (\ref{fock.norm1}) is the most convenient way to write things.

So now given a function $F\in\mathcal{H}L^{2}(\mathbb{C}^{d},\mu_{t}),$ we
wish to produce an element of the Fock space $\xi.$ We do this by setting
\begin{equation}
\alpha_{k_{1},\cdots,k_{n}}=\left(  \frac{\partial^{n}}{\partial z_{k_{1}%
}\cdots\partial z_{k_{n}}}F\right)  \left(  0\right)  .\label{taylor1}%
\end{equation}
That is, the $\alpha$'s are the Taylor coefficients of $F$ at the origin. Note
that the $\alpha$'s defined in this way are symmetric. Using a Taylor
expansion of $F,$ it is not difficult to see that
\[
\left\|  F\right\|  _{L^{2}\left(  \mathbb{C}^{d},\mu_{t}\right)  }^{2}%
=\sum_{n=0}^{\infty}\frac{t^{n}}{n!}\sum_{k_{1},\cdots,k_{n}}\left|
\alpha_{k_{1},\cdots,k_{n}}\right|  ^{2}.
\]
In the case $d=1$ this amounts to the statement that the functions $\left\{
z^{n}\right\}  _{n=0}^{\infty}$ form an orthogonal basis of $\mathcal{H}%
L^{2}\left(  \mathbb{C},\mu_{t}\right)  .$ Thus the ``Taylor map'' that
associates to $F$ the collection of coefficients $\left\{  \alpha
_{k_{1},\cdots,k_{n}}\right\}  $ (as in (\ref{taylor1})) is an isometry of
$\mathcal{H}L^{2}(\mathbb{C}^{d},\mu_{t})$ into the Fock space $I_{t}%
^{0}(\mathbb{R}^{d}).$ It is easily seen that the Taylor map is \textit{onto}
the Fock space. Inverting the Taylor map amounts to expanding $F$ in terms of
the monomials $z_{1}^{n_{1}}\cdots z_{d}^{n_{d}}$---a Taylor series.

Finally, we consider the \textbf{Hermite expansion}, mapping from the position
Hilbert space $L^{2}(\mathbb{R}^{d},\rho_{t})$ to the Fock space. This map is
the composition of the Segal-Bargmann transform and the Taylor map. Since $F$
is holomorphic, its derivatives at zero may be computed by differentiating
just in the real directions. Thus the Taylor coefficients of $B_{t}f$ (see
(\ref{bt.form1}) and (\ref{taylor1})) may be expressed as
\begin{equation}
\alpha_{k_{1},\cdots,k_{n}}=\left(  \frac{\partial^{n}}{\partial x_{k_{1}%
}\cdots\partial x_{k_{n}}}e^{t\Delta/2}f\right)  \left(  0\right)
.\label{herm.form}%
\end{equation}
So we apply the heat equation to $f,$ then compute all derivatives at the
origin. In light of the two previous theorems, the map $f\rightarrow\left\{
\alpha_{k_{1},\cdots,k_{n}}\right\}  $ given in (\ref{herm.form}) is an
isometry of $L^{2}(\mathbb{R}^{d},\rho_{t})$ onto the Fock space. Inverting
this map amounts to expanding $f$ in terms of the functions $H_{n_{1}%
,\cdots,n_{d}}$ such that
\[
B_{t}\left(  H_{n_{1},\cdots,n_{d}}\right)  =z_{1}^{n_{1}}\cdots z_{d}^{n_{d}%
}.
\]
These functions are the \textbf{Hermite polynomials} given by
\[
H_{n_{1},\cdots,n_{d}}=e^{-t\Delta/2}\left(  x_{1}^{n_{1}}\cdots x_{d}^{n_{d}%
}\right)  .
\]
Here $e^{-t\Delta/2}$ is the \textit{inverse} heat operator, which is computed
as a terminating power series. These polynomials may also be expressed in
terms of derivatives of a Gaussian, as in Section 2.4.

Of course, there are many important aspects of these three Hilbert spaces that
I have not discussed here. Most importantly, one should consider the natural
operators on these spaces, namely, the creation and annihilation operators,
and operators constructed from them. I discuss these operators in the group
case in Section 2.5, with some comments about the simplifications that occur
in the $\mathbb{R}^{d}$ case.

It is a pleasure to thank those who have made corrections to this article:
Maria Gordina, Wicharn Lewkeeratiyutkul, Jeffrey Mitchell, Stephen Sontz, and
Matthew Stenzel.

\section{Three unitary maps for a compact Lie group}

Additional exposition of these results can be found in \cite{G9}.

\subsection{Preliminaries}

Let $K$ be a connected Lie group of \textbf{compact type}. A Lie group is said
to be of compact type if it is locally isomorphic to some compact Lie group.
Equivalently, a Lie group $K$ is of compact type if there exists an inner
product on the Lie algebra of $K$ that is invariant under the adjoint action
of $K$. So $\mathbb{R}^{d}$ is of compact type, being locally isomorphic to a
$d$-torus, and every compact Lie group is of compact type. It can be shown
that every connected Lie group of compact type is isomorphic to a product of
$\mathbb{R}^{d}$ and a connected compact Lie group. So all of the
constructions described here for Lie groups of compact type include as a
special case the constructions for $\mathbb{R}^{d}.$ On the other hand, all
the new information (beyond the $\mathbb{R}^{d}$ case) is contained in the
compact case. See \cite[Chap. II, Sect. 6]{He} (including Proposition 6.8) for
information on Lie groups of compact type.

Let $\frak{k}$ denote the Lie algebra of $K.$ We fix once and for all an inner
product $\left\langle \cdot,\cdot\right\rangle $ on $\frak{k}$ that is
invariant under the adjoint action of $K.$ For example we may take
$K=\mathrm{SU}(n),$ in which case $\frak{k}=\mathrm{su}(n)$ is the space of
skew matrices with trace zero. An invariant inner product on $\frak{k}$ is
$\left\langle X,Y\right\rangle =\operatorname{Re}\left[  \mathrm{trace}\left(
X^{\ast}Y\right)  \right]  .$

Now let $K_{\mathbb{C}}$ be the \textbf{complexification} of $K$. If $K$ is
simply connected then the complexification of $K$ is the unique simply
connected Lie group whose Lie algebra $\frak{k}_{\mathbb{C}}$ is
$\frak{k}+i\frak{k}.$ In general, $K_{\mathbb{C}}$ is defined by the following
three properties. First, $K_{\mathbb{C}}$ should be a connected complex Lie
group whose Lie algebra $\frak{k}_{\mathbb{C}}$ is equal to $\frak{k}%
+i\frak{k}.$ Second, $K_{\mathbb{C}}$ should contain $K$ as a closed subgroup
(whose Lie algebra is $\frak{k}\subset\frak{k}_{\mathbb{C}}$). Third, every
homomorphism of $K$ into a complex Lie group $H$ should extend to a
holomorphic homomorphism of $K_{\mathbb{C}}$ into $H.$ The complexification of
a connected Lie group of compact type always exists and is unique.

\begin{example}
If $K=\mathbb{R}^{d}$ then $K_{\mathbb{C}}=\mathbb{C}^{d}.$ If $K=\mathrm{SU}%
(n)$ then $K_{\mathbb{C}}=\mathrm{SL}(n;\mathbb{C}).$ If $K=\mathrm{SO}(n)$
then $K_{\mathbb{C}}=\mathrm{SO}(n;\mathbb{C}).$ In the first two examples,
$K$ and $K_{\mathbb{C}}$ are simply connected. In the last example, neither
$K$ nor $K_{\mathbb{C}}$ is simply connected.
\end{example}

We have the following structure theorem for Lie groups of compact type. This
result is a modest strengthening of Corollary 2.2 of \cite{Dr} and allows all
the relevant results for Lie groups of compact type to be reduced to two
cases, the compact case and the $\mathbb{R}^{d}$ case.

\begin{proposition}
\label{structure.prop}Suppose that $K$ is a connected Lie group of compact
type, with a fixed Ad-invariant inner product on its Lie algebra $\frak{k}.$
Then there exists a decomposition $K=K_{1}\times K_{2},$ where $K_{1}$ is
compact and $K_{2}$ is isomorphic to $\mathbb{R}^{d},$ and such that the
associated Lie algebra decomposition $\frak{k}=\frak{k}_{1}+\frak{k}_{2}$ is orthogonal.
\end{proposition}

\subsection{The Segal-Bargmann transform for $K$}

The Ad-$K$-invariant inner product on $\frak{k}$ determines a bi-invariant
Riemannian metric on $K.$ Let $\Delta_{K}$ be the (non-positive) Laplacian
associated to this metric. Let $dx$ denote Haar measure on $K$, normalized to
coincide with Riemannian volume measure. Then let $\rho_{t}$ denote the
\textbf{heat kernel} (at the identity) on $K,$ i.e. the solution to
\[
\frac{d\rho}{dt}=\frac{1}{2}\Delta_{K}\rho_{t}
\]
subject to the initial condition
\[
\lim_{t\downarrow0}\int_{K}f\left(  x\right)  \rho_{t}\left(  x\right)
\,dx=f\left(  e\right)
\]
for all continuous functions $f$ of compact support. We also let $\rho_{t}$
denote the \textbf{heat kernel measure}
\[
d\rho_{t}\left(  x\right)  :=\rho_{t}\left(  x\right)  \,dx.
\]
If $K=\mathbb{R}^{d}$ (with the standard metric) then $\rho_{t}$ is nothing
but the Gaussian measure (\ref{gauss.def1}). The following result is proved in
\cite[Sect. 2]{H1}.

\begin{proposition}
For all $t>0$ the function $\rho_{t}$ has a unique analytic continuation from
$K$ to $K_{\mathbb{C}}.$
\end{proposition}

Thus we may define the \textbf{Segal-Bargmann transform} for $K$ by analogy to
the $\mathbb{R}^{d}$ case.

\begin{definition}
\label{sb.def2}Let $\mathcal{H}(K_{\mathbb{C}})$ denote the space of
holomorphic functions on $K_{\mathbb{C}}.$ Then for each $t>0$ define a map
$B_{t}:L^{2}(K,\rho_{t})\rightarrow\mathcal{H}(K_{\mathbb{C}})$ by
\[
B_{t}f\left(  g\right)  =\int_{K}\rho_{t}\left(  gx^{-1}\right)  f\left(
x\right)  \,dx,\quad g\in K_{\mathbb{C}},
\]
where on the right $\rho_{t}\left(  gx^{-1}\right)  $ refers to the
analytically continued heat kernel.
\end{definition}

It is not obvious but true that the integral defining $B_{t}f$ is always
convergent, and that the result is holomorphic as a function of $g.$ In the
compact case there is no problem, and in the $\mathbb{R}^{d}$ case it is a
straightforward calculation.

Note that the restriction of $B_{t}f$ to $K$ is just the convolution of
$\rho_{t}$ with $f.$ Since $\rho_{t}$ is the heat kernel, this means that the
restriction of $B_{t}f$ to $K$ satisfies the heat equation, with initial
condition $f.$ Thus we may alternatively write
\begin{equation}
B_{t}f=\text{ analytic continuation of }e^{t\Delta_{K}/2}f,\label{bt.form3}%
\end{equation}
where $e^{t\Delta_{K}/2}f$ is shorthand for the solution of the heat equation
given by the convolution of $\rho_{t}$ with $f.$ The analytic continuation is
in the space variable, from $K$ to $K_{\mathbb{C}},$ with the time parameter
$t$ fixed.

We now construct an appropriate heat kernel measure on $K_{\mathbb{C}}.$ We
extend the inner product on $\frak{k}$ to a real-valued inner product on
$\frak{k}_{\mathbb{C}}$ given by
\[
\left\langle X_{1}+iY_{1},X_{2}+iY_{2}\right\rangle =\left\langle X_{1}%
,X_{2}\right\rangle +\left\langle Y_{1},Y_{2}\right\rangle .
\]
Then there is a unique left-invariant Riemannian metric on $K_{\mathbb{C}}$
given by this inner product at the identity. Let $\Delta_{K_{\mathbb{C}}}$ be
the Laplacian for this metric. We now define $\mu_{t}$ to be the \textbf{heat
kernel} (at the identity) on $K_{\mathbb{C}},$ that is, the solution to
\[
\frac{d\mu}{dt}=\frac{1}{4}\Delta_{K_{\mathbb{C}}}\mu_{t}
\]
subject to the initial condition
\[
\lim_{t\downarrow0}\int_{K_{\mathbb{C}}}f\left(  g\right)  \mu_{t}\left(
g\right)  \,dg=f\left(  e\right)
\]
for all continuous functions $f$ with compact support. Here $dg$ is the
(bi-invariant) Haar measure on $K_{\mathbb{C}}$, normalized to coincide with
Riemannian volume measure. We also let $\mu_{t}$ denote the associated
\textbf{heat kernel measure}
\[
d\mu_{t}\left(  g\right)  :=\mu_{t}\left(  g\right)  \,dg.
\]

We are now ready to state the main theorem about the Segal-Bargmann transform
for $K.$

\begin{theorem}
\label{sb.thm2}For all $t>0$ the map $B_{t}$ given in Definition \ref{sb.def2}
is a unitary map of $L^{2}(K,\rho_{t})$ onto $\mathcal{H}L^{2}(K_{\mathbb{C}%
},\mu_{t}),$ where $\mathcal{H}L^{2}(K_{\mathbb{C}},\mu_{t})$ denotes the
space of holomorphic functions on $K_{\mathbb{C}}$ that are square-integrable
with respect to $\mu_{t}.$
\end{theorem}

This is Theorem 1$^{\prime}$ of \cite{H1}. (More precisely, \cite{H1} proves
the compact case. The case of general compact-type groups can be reduced to
the compact case and the $\mathbb{R}^{d}$ case due to Segal and Bargmann.) A
sketch of the proof of this result and the other main theorems is given in
Section 5.

\subsection{The Taylor map}

Let $\left\{  X_{k}\right\}  _{k=1}^{\dim\frak{k}}$ be an orthonormal basis
for $\frak{k}.$ We think of each $X_{k}$ as a left-invariant vector field on
$K_{\mathbb{C}},$ satisfying
\[
X_{k}f\left(  g\right)  =\left.  \frac{d}{ds}\right|  _{s=0}f\left(
ge^{sX_{k}}\right)  ,\quad g\in K_{\mathbb{C}}.
\]
Then the norm of a function $F\in\mathcal{H}L^{2}\left(  K_{\mathbb{C}}%
,\mu_{t}\right)  $ can be expressed as follows in terms of its derivatives at
the identity in the following way.

\begin{theorem}
\label{taylor.thm2}Let $K$ be a connected Lie group of compact type, and
$K_{\mathbb{C}}$ the complexification of $K.$ Let $\left\{  X_{k}\right\}
_{k=1}^{\dim\frak{k}}$ be an orthonormal basis for $\frak{k}$ (with respect to
a fixed Ad-$K$-invariant inner product). Then for all $F\in\mathcal{H}%
L^{2}(K_{\mathbb{C}},\mu_{t})$%
\[
\int_{K_{\mathbb{C}}}\left|  F\left(  g\right)  \right|  ^{2}\mu_{t}\left(
g\right)  \,dg=\sum_{n=0}^{\infty}\frac{t^{n}}{n!}\sum_{k_{1},\cdots,k_{n}%
=1}^{\dim\frak{k}}\left|  \left(  X_{k_{1}}X_{k_{2}}\cdots X_{k_{n}}F\right)
\left(  e\right)  \right|  ^{2}.
\]
\end{theorem}

This Theorem and the succeeding one are consequences of Theorem 1$^{\prime}$
of \cite{H1} and Proposition 2.4 of \cite{Hi1}. They are stated explicitly in
\cite[Cor. 1.17]{Dr}. At a formal level the above identity is easily obtained
by expanding the left side in powers of $t$. (See Section 5.2.) This result
expresses the norm of $F$ in terms of its ``Taylor coefficients,'' namely, the
left-invariant derivatives of $F$ at the identity.

We wish to understand what possible collections of Taylor coefficients can
arise, so as to have an algebraic description of our Hilbert space, comparable
to the Fock symmetric tensor model in the $\mathbb{R}^{d}$ case. There are two
restriction on the Taylor coefficients. First there is an algebraic condition:
the commutation relations of the Lie algebra impose certain linear relations
among the derivatives $\left(  X_{k_{1}}X_{k_{2}}\cdots X_{k_{n}}F\right)
\left(  e\right)  .$ For example, $X_{j}X_{k}$ is equal to $X_{k}X_{j}$ plus
$\left[  X_{j},X_{k}\right]  ,$ which is a linear combination of the $X_{l}%
$'s. So a certain linear combination of $X_{j}X_{k}$, $X_{k}X_{j},$ and the
$X_{l}$'s is zero, and this linear relation is necessarily reflected in the
derivatives of $F$ at $e.$ Similar relations exist among higher-order
derivatives. Second there is an analytic condition: the weighted sum of the
squares of the derivatives, as in Theorem \ref{taylor.thm2}, must be finite.
If $K$ (or equivalently $K_{\mathbb{C}}$) is simply connected, these two
conditions are sufficient.

\begin{theorem}
Suppose $\left\{  \alpha_{k_{1},\cdots,k_{n}}\right\}  $ is a collection of
complex numbers satisfying the algebraic relations associated to the Lie
algebra $\frak{k}$ and such that
\[
\sum_{n=0}^{\infty}\frac{t^{n}}{n!}\sum_{k_{1},\cdots,k_{n}=1}^{\dim\frak{k}%
}\left|  \alpha_{k_{1},\cdots,k_{n}}\right|  ^{2}<\infty.
\]
Suppose also that $K$ is simply connected. Then there exists a unique
$F\in\mathcal{H}L^{2}(K_{\mathbb{C}},\mu_{t})$ such that $\left(  X_{k_{1}%
}X_{k_{2}}\cdots X_{k_{n}}F\right)  \left(  e\right)  =\alpha_{k_{1}%
,\cdots,k_{n}}.$
\end{theorem}

The basis-independent way of saying this is as follows. Consider the tensor
algebra over $\frak{k},$ $T(\frak{k}).$ Let $J$ be the two-sided ideal in
$T(\frak{k})$ generated by elements of the form
\[
X\otimes Y-Y\otimes X-\left[  X,Y\right]  ,\quad X,Y\in\frak{k}.
\]
Then the quotient algebra $U\left(  \frak{k}\right)  :=T(\frak{k})/J$ is the
\textbf{universal enveloping algebra} of $\frak{k}.$ Let $T(\frak{k})^{\ast} $
be the complex dual space, i.e. the set of all linear maps of $T(\frak{k})$
into $\mathbb{C}.$ This may be expressed as
\[
\sum_{n=0}^{\infty}(\frak{k}^{\ast})^{\otimes n}\quad\text{(strong direct
sum),}
\]
where $\frak{k}^{\ast}$ is the complex dual space of $\frak{k}.$ There is a
natural norm $\left\|  \cdot\right\|  _{n}$ on ($\frak{k}^{\ast})^{\otimes n}
$ given as in (\ref{xin.norm}) above.

Now a general element $\alpha$ of $T(\frak{k})^{\ast}$ decomposes as $\xi
=\xi_{0}+\xi_{1}+\xi_{2}+\cdots,$ with $\xi_{n}\in(\frak{k}^{\ast})^{\otimes
n}.$ Define a ($t$-dependent) norm $\left\|  \cdot\right\|  _{t}$ on
$T(\frak{k})^{\ast}$ by
\[
\left\|  \xi\right\|  _{t}^{2}=\sum_{n=0}^{\infty}\frac{t^{n}}{n!}\left\|
\xi_{n}\right\|  _{n}^{2}.
\]
Since we impose no condition on the sequence $\xi_{n}\in\left(  \frak{k}%
^{\ast}\right)  ^{\otimes n}$, the norm $\left\|  \xi\right\|  _{t}$ may take
the value $+\infty.$

\begin{definition}
Let $J^{0}$ be the set of linear functionals $\xi$ in $T(\frak{k})^{\ast}$
such that $\xi$ is zero on the ideal $J.$ Equivalently, elements $\xi\in
J^{0}$ may be thought of as linear functionals on the universal enveloping
algebra $T(\frak{k})/J.$

Let $J_{t}^{0}$ be the set of elements $\xi\in J^{0}$ such that $\left\|
\xi\right\|  _{t}<\infty.$
\end{definition}

Now as usual we may think of the universal enveloping algebra $U(\frak{k}%
)=T(\frak{k})/J$ as the algebra of left-invariant differential operators on
$K.$ Given a function $F\in\mathcal{H}L^{2}(K_{\mathbb{C}},\mu_{t})$ we wish
to define a linear functional on $U(\frak{k})$---that is, an element of
$J^{0}$---called the \textbf{Taylor map of }$F.$ This is given by
\[
\text{Taylor}\left(  F\right)  \left(  \alpha\right)  =\left(  \alpha
F\right)  \left(  e\right)  ,\quad\alpha\in U(\frak{k}).
\]
On the right we are (implicitly) restricting $F$ to $K,$ then applying the
differential operator $\alpha,$ then evaluating at the identity. One could
more generally apply left-invariant derivatives on $K_{\mathbb{C}}$ to $F$ and
evaluate at the identity. But since $F$ is holomorphic, these could all be
reduced to derivatives in the $K$ directions.

We may now rephrase the previous two results in basis-independent terms as follows.

\begin{theorem}
The Taylor map $F\rightarrow$Taylor$\left(  F\right)  $ is an isometric map of
$\mathcal{H}L^{2}(K_{\mathbb{C}},\mu_{t})$ into $J_{t}^{0}.$ If $K$ is simply
connected then the Taylor map takes $\mathcal{H}L^{2}(K_{\mathbb{C}},\mu_{t})$
\textbf{onto} $J_{t}^{0}.$
\end{theorem}

The proof of isometricity of the Segal-Bargmann transform relies on the fact
that inner product on $\frak{k}$ is Ad-$K$-invariant. This is why the real
group $K$ must be of compact type (i.e. admit an Ad-invariant inner product).
But an examination of the proof of the isometricity of the Taylor map (say as
given in \cite{Dr}) shows that this does not use the Ad-invariance of the
inner product. In fact, as shown by Driver and Gross, this isomorphism can be
generalized to arbitrary complex Lie groups, as I will now describe. I revert
at this stage to basis-dependent notation.

Let $G$ be an arbitrary connected complex Lie group with Lie algebra
$\frak{g}$ and complex dimension $d.$ Let $J:\frak{g}\rightarrow\frak{g}$
denote the complex structure on $\frak{g}.$ Fix an arbitrary Hermitian inner
product $\left(  \cdot,\cdot\right)  $ on $\frak{g}.$ Then $\left\langle
\cdot,\cdot\right\rangle :=\operatorname{Re}\left(  \cdot,\cdot\right)  $ is a
real-valued inner product on $\frak{g},$ which is invariant under $J.$ Let
$X_{1},\cdots,X_{d}$ be a basis for $\frak{g}$ as a complex vector space that
is orthonormal with respect to $\left(  \cdot,\cdot\right)  .$ Then
$X_{1},\cdots,X_{d},JX_{1},\cdots,JX_{d}$ is an orthonormal basis for
$\frak{g}$ as a real vector space. We regard $X_{1},\cdots,X_{d},JX_{1}%
,\cdots,JX_{d}$ as left-invariant vector fields on $G$. For a general smooth
function $f,$ $\left(  JX\right)  f$ is not the same as $i\left(  Xf\right)
,$ although these two quantities are equal if $f$ is holomorphic. Now define
\[
\Delta_{G}=\sum_{k=1}^{d}\left[  X_{k}^{2}+\left(  JX_{k}\right)  ^{2}\right]
.
\]
We call this the Laplacian for $G.$ If $G$ is unimodular, then $\Delta_{G}$
coincides with the Laplace-Beltrami operator for $G$ with respect to the
left-invariant Riemannian metric determined by $\left\langle \cdot
,\cdot\right\rangle .$

Let $\mu_{t}$ denote the ``heat kernel'' for $\Delta_{G},$ i.e. the solution
of
\[
\frac{d\mu}{dt}=\frac{1}{4}\Delta_{G}\mu_{t}\left(  g\right)
\]
subject to the initial condition
\[
\lim_{t\downarrow0}\int_{G}f\left(  g\right)  \mu_{t}\left(  g\right)
\,dg=f\left(  e\right)  ,
\]
for all continuous functions $f$ with compact support, where $dg$ is a fixed
\textit{right-invariant} Haar measure on $G.$ The use of a left-invariant
Laplacian and a right-invariant Haar measure is so that $\Delta_{G}$ will be
self-adjoint in $L^{2}\left(  G,dg\right)  .$

\begin{theorem}
\label{dg.thm}If $F\in\mathcal{H}L^{2}(G,\mu_{t}\left(  g\right)  dg)$ then
\[
\int_{G}\left|  F\left(  g\right)  \right|  ^{2}\mu_{t}\left(  g\right)
\,dg=\sum_{n=0}^{\infty}\frac{t^{n}}{n!}\sum_{k_{1},\cdots,k_{n}=1}^{d}\left|
\left(  X_{k_{1}}X_{k_{2}}\cdots X_{k_{n}}F\right)  \left(  e\right)  \right|
^{2}.
\]
Suppose that $G$ is simply connected. Suppose also that $\alpha_{k_{1}%
,\cdots,k_{n}}$ is a collection of complex numbers satisfying the algebraic
relations associated to the Lie algebra $\frak{g}$ and such that
\[
\sum_{n=0}^{\infty}\frac{t^{n}}{n!}\sum_{k_{1},\cdots,k_{n}=1}^{d}\left|
\alpha_{k_{1},\cdots,k_{n}}\right|  ^{2}<\infty.
\]
Then there exists a unique $F\in\mathcal{H}L^{2}(G,\mu_{t}\left(  g\right)
dg)$ such that $\left(  X_{k_{1}}X_{k_{2}}\cdots X_{k_{n}}F\right)  \left(
e\right)  =\alpha_{k_{1},\cdots,k_{n}}.$
\end{theorem}

These results are Theorems 2.5 and 2.6 of \cite{DG}. M. Gordina has obtained
similar results for certain \textit{infinite-dimensional} complex Lie groups
\cite{Go1,Go2}.

The Taylor expansion gives the following useful pointwise bound on functions
in $\mathcal{H}L^{2}\left(  G,\mu_{t}\left(  g\right)  \,dg\right)  .$

\begin{theorem}
\label{bounds.thm}Let $G$ and $\mu_{t}$ be as in Theorem \ref{dg.thm}. For
$g\in G,$ let $\left|  g\right|  $ denote the distance from the identity to
$g$ with respect to the left-invariant Riemannian metric on $G$ determined by
the inner product $\left\langle \cdot,\cdot\right\rangle $ on $\frak{g}.$ Then
for all $F\in$ $\mathcal{H}L^{2}(G,\mu_{t}\left(  g\right)  dg)$ we have
\[
\left|  F\left(  g\right)  \right|  ^{2}\leq\left\|  F\right\|  _{L^{2}\left(
G,\mu_{t}\right)  }^{2}e^{\left|  g\right|  ^{2}/t}.
\]
\end{theorem}

This result follows from Corollary 3.10 and Remark 3.11 of \cite{DG}, with
$k=0.$ In the case $G=K_{\mathbb{C}}$ (with an Ad-$K$-invariant inner product
on $\frak{k}_{\mathbb{C}}$), this result was obtained previously in \cite[Cor.
5.5 and Thm. 5.7]{Dr}. (Compare \cite[Eq. (1.7)]{B}.) In the case
$G=K_{\mathbb{C}}$ it is possible to give exponentially better bounds than
those in Theorem \ref{bounds.thm}. (See Section 3.3 and \cite{H3}.)
Nevertheless the bounds in Theorem \ref{bounds.thm} are very useful and play
an important role in the proofs in \cite{Dr} and \cite{DG}, as well as
\cite{HS,Go1,Go2}.

\begin{problem}
Suppose that $G,$ $\mu_{t}$ are as in Theorem \ref{dg.thm}. Assuming that $G $
has a faithful finite-dimensional representation, prove that the matrix
entries of the finite-dimensional holomorphic representations span a dense
subspace of $\mathcal{H}L^{2}(G,\mu_{t}\left(  g\right)  dg).$
\end{problem}

\begin{problem}
Suppose that $G,$ $\mu_{t}$ are as in Theorem \ref{dg.thm}. Prove that for all
$\varepsilon>0,$ $\mathcal{H}L^{2}(G,\mu_{t+\varepsilon}\left(  g\right)  dg)$
is dense in $\mathcal{H}L^{2}(G,\mu_{t}\left(  g\right)  dg).$
\end{problem}

If $G=K_{\mathbb{C}}$ and the inner product on $\frak{g}$ is invariant under
the adjoint action of $K,$ then both problems have been solved in the
affirmative, as a consequence of the ``averaging lemma'' \cite[Lem. 11]{H1}.
But even the case $G=K_{\mathbb{C}}$ with an arbitrary Hermitian inner product
on $\frak{g}$ is open.

\subsection{The Hermite expansion}

Note that since each vector field $X_{k}$ in Theorem \ref{taylor.thm2} is in
$\frak{k},$ all the derivatives on the right in that theorem involve only the
values of $F$ on $K.$ (Since $F$ is holomorphic, all of the derivatives at the
identity, in any direction, can be computed in terms of derivatives in the
$K$-directions.) Suppose now that $F$ is the Segal-Bargmann transform of some
function $f$ in $L^{2}(K,\rho_{t}).$ Then the restriction of $F$ to $K$ is
just $e^{t\Delta_{K}/2}f$ (see (\ref{bt.form3})). So combining the
isometricity of the Segal-Bargmann transform and the isometricity of the
Taylor map we get the following result.

\begin{theorem}
\label{hermite.thm2}Let $K$ be a connected Lie group of compact type, and
$\left\{  X_{k}\right\}  _{k=1}^{\dim\frak{k}}$ an orthonormal basis for
$\frak{k}$ with respect to a fixed Ad-$K$-invariant inner product. Then for
all $f\in L^{2}(K,\rho_{t})$
\[
\int_{K}\left|  f\left(  x\right)  \right|  ^{2}\rho_{t}\left(  x\right)
\,dx=\sum_{n=0}^{\infty}\frac{t^{n}}{n!}\sum_{k_{1},\cdots,k_{n}=1}%
^{\dim\frak{k}}\left|  \left(  X_{k_{1}}X_{k_{2}}\cdots X_{k_{n}}%
e^{t\Delta_{K}/2}f\right)  \left(  e\right)  \right|  ^{2}.
\]

Suppose that $\left\{  \alpha_{k_{1},\cdots,k_{n}}\right\}  $ is a collection
of complex numbers satisfying the algebraic relations associated to the Lie
algebra $\frak{k}$ and such that
\[
\sum_{n=0}^{\infty}\frac{t^{n}}{n!}\sum_{k_{1},\cdots,k_{n}=1}^{\dim\frak{k}%
}\left|  \alpha_{k_{1},\cdots,k_{n}}\right|  ^{2}<\infty.
\]
Suppose also that $K$ is simply connected. Then there exists a unique $f\in
L^{2}(K,\rho_{t})$ such that $\left(  X_{k_{1}}X_{k_{2}}\cdots X_{k_{n}%
}e^{t\Delta_{K}/2}f\right)  \left(  e\right)  =\alpha_{k_{1},\cdots,k_{n}}$
for all $k_{1},\cdots,k_{n}.$
\end{theorem}

This result is a consequence of \cite[Thm. 2.1]{G2}. (See also \cite{G3,G4}.)
The result as given here first appears in \cite[Props. 2.4 and 2.5]{Hi1}. See
also \cite[Thm. 1.4]{Dr} and \cite{Hi2}.

The Hermite expansion leads naturally to a notion of \textbf{Hermite
functions} on $K,$ namely the functions $H_{k_{1},\cdots,k_{n}}$ on $K$ such
that
\[
\left(  X_{k_{1}}X_{k_{2}}\cdots X_{k_{n}}e^{t\Delta_{K}/2}f\right)  \left(
e\right)  =\left\langle H_{k_{1},\cdots,k_{n}},f\right\rangle _{L^{2}\left(
K,\rho_{t}\right)  }.
\]
A straightforward calculation establishes the following Rodriguez-type formula
for the Hermite functions
\[
H_{k_{1},\cdots,k_{n}}=\left(  -1\right)  ^{n}\frac{X_{k_{n}}\cdots X_{k_{1}%
}\rho_{t}}{\rho_{t}}.
\]
In the case $K=\mathbb{R}^{d}$ these are the usual Hermite polynomials. In the
compact case these functions have been studied by J. Mitchell. In \cite{M1} he
shows that after a suitable re-scaling, the generalized Hermite polynomials
converge as $t$ tends to zero to the ordinary Hermite polynomials. In
\cite{M2} he develops a small-$t$ asymptotic expansion for the Hermite
functions, in terms of ordinary Hermite polynomials.

\subsection{Creation and annihilation operators}

Gross originally \cite{G2} described the Hermite expansion in terms of the
action of the relevant ``annihilation operators.'' In this section we will
consider the \textbf{annihilation operators} in each of our three unitarily
related Hilbert spaces, as well as their adjoints, the \textbf{creation
operators}. The creation and annihilation operators are also called raising
and lowering operators.

Starting with a connected Lie group $K$ of compact type, we have considered
three Hilbert spaces: the position Hilbert space $L^{2}(K,\rho_{t});$ the
Segal-Bargmann space $\mathcal{H}L^{2}(K_{\mathbb{C}},\mu_{t});$ and the
Fock-type ``tensor'' space $J_{t}^{0}$ (the space of linear functionals on
$T(\frak{k})/J$ with finite norm). Each of these three spaces has a naturally
defined set of creation and annihilation operators, labeled by elements of the
Lie algebra $\frak{k}.$ At some risk of confusion, I am going to use the same
symbols for the creation and annihilation operators in each of the three
Hilbert spaces. So in each Hilbert space we will have a family of annihilation
operators $a_{X},$ labeled by elements $X$ of the Lie algebra $\frak{k},$ and
also a corresponding family of creation operators $a_{X}^{\ast}.$ In each case
the creation operators are defined to be the adjoints of the corresponding
annihilation operators: $a_{X}^{\ast}:=\left(  a_{X}\right)  ^{\ast}$ (in case
this is not already evident from the notation). In addition to the creation
and annihilation operators there is a \textbf{vacuum state} (or ground state),
a unique (up to a constant) element of the Hilbert space that is annihilated
by all the annihilation operators. In each case the annihilation operators
satisfy the commutation relations of the Lie algebra: $\left[  a_{X}%
,a_{Y}\right]  =a_{\left[  X,Y\right]  }.$ As a consequence the creation
operators satisfy $\left[  a_{X}^{\ast},a_{Y}^{\ast}\right]  =-a_{\left[
X,Y\right]  }^{\ast}.$

\textit{The position Hilbert space }$L^{2}\left(  K,\rho_{t}\right)  .$ We
think of the Lie algebra elements $X$ as left-invariant vector fields. We
define the annihilation operators to be simply these vector fields:
\[
a_{X}f\left(  x\right)  =Xf\left(  x\right)  =\left.  \frac{d}{ds}\right|
_{s=0}f\left(  xe^{sX}\right)  .
\]
The vacuum state is the constant function $\mathbf{1}$. We then define the
creation operators to be the adjoints of the left-invariant vector fields,
computed with respect to the inner product for $L^{2}(K,\rho_{t}).$ Explicitly
we have
\[
a_{X}^{\ast}=-X-X\left(  \log\rho_{t}\right)  .
\]
Here the second term is a multiplication operator, by the derivative of the
logarithm of the heat kernel. In the $\mathbb{R}^{d}$ case, $X(\log\rho_{t})$
is just a linear function.

\textit{The Segal-Bargmann space }$\mathcal{H}L^{2}(K_{\mathbb{C}},\mu_{t}).$
Here the annihilation operators are again defined to be left-invariant vector
fields, but now regarded as acting on $K_{\mathbb{C}}$ instead of $K.$ To
avoid confusion, we now let $J:\frak{k}_{\mathbb{C}}\rightarrow\frak{k}%
_{\mathbb{C}}$ denote the complex structure on $\frak{k}_{\mathbb{C}}.$ For a
general smooth function $f$ on $K_{\mathbb{C}},$ $(JX)f\neq i(Xf).$ Indeed,
$f$ is holomorphic if and only if $\left(  JX\right)  f=i\left(  Xf\right)  $
for all $X\in\frak{k}_{\mathbb{C}}.$

When acting on holomorphic functions the real vector field $X$ coincides with
the holomorphic vector field $\left(  X-iJX\right)  /2.$ So in the
Segal-Bargmann space we have
\begin{align*}
a_{X}F\left(  g\right)   & =\frac{1}{2}\left(  X-iJX\right)  F\left(  g\right)
\\
& =\frac{1}{2}\frac{d}{ds}\left.  \left(  F\left(  ge^{sX}\right)  -iF\left(
ge^{sJX}\right)  \right)  \right|  _{s=0}.
\end{align*}
These operators preserve the space of holomorphic functions. In the case
$K_{\mathbb{C}}=\mathbb{C}^{d}$ these holomorphic vector fields are just
linear combinations of the operators $\partial/\partial z_{k}:=(\partial
/\partial x_{k}-i\partial/\partial y_{k})/2.$ In all cases the vacuum state is
the constant function $\mathbf{1}.$

The creation operators (defined to be the adjoints of the annihilation
operators) can be computed as \textbf{Toeplitz operators}, namely,
\[
a_{X}^{\ast}F=P_{t}\left(  \phi_{X,t}F\right)
\]
where
\[
\phi_{X,t}=-\frac{1}{2}(X+iJX)(\log\mu_{t}).
\]
Here $P_{t}$ is the orthogonal projection operator from $L^{2}(K_{\mathbb{C}%
},\mu_{t})$ onto the holomorphic subspace $\mathcal{H}L^{2}(K_{\mathbb{C}}%
,\mu_{t}).$ So we take the holomorphic function $F,$ multiply by the typically
non-holomorphic function $-(X+iJX)(\log\mu_{t})/2,$ and then project back into
the holomorphic subspace. In verifying this formula for $a_{X}^{\ast}$ one
uses integration by parts and the Cauchy-Riemann equations for $F$ in the form $(JX)F=i(XF).$

In the case $K_{\mathbb{C}}=\mathbb{C}^{d},$ the operators $(1/2)(X+iJX)$ are
just linear combinations of the operators $\partial/\partial\bar{z}_{k}.$ We
compute that $-\partial\log\mu_{t}/\partial\bar{z}_{k}$ $=z_{k}/t.$ Since in
this case the logarithmic derivative is holomorphic, the projection is
unnecessary and the creation operators is just multiplication by $z_{k}/t. $

\textit{The ``Fock space'' }$J_{t}^{0}.$ Recall that elements of $J_{t}^{0}$
are linear functionals on the universal enveloping algebra $T(\frak{k})/J$
(with finite norm). The annihilation operators in $J_{t}^{0}$ are defined as
follows. Given $X\in\frak{k}$ and $\xi\in J_{t}^{0},$ we define
\[
a_{X}\xi(\alpha)=\xi(\alpha X),\quad\alpha\in T(\frak{k})/J.
\]
That is, $a_{X}$ is the adjoint of right-multiplication by $X$ in the
universal enveloping algebra $T(\frak{k})/J.$ The vacuum state in $J_{t}^{0}$
is the linear functional $\phi_{0}$ that picks out the coefficient of the
identity for each $\alpha\in T(\frak{k})/J.$ The creation operators are the
adjoints of the annihilation operators, computed with respect to the natural
($t$-dependent) inner product on $J_{t}^{0}.$ If one uses the inner product to
identify $J_{t}^{0}$ with a completion of the enveloping algebra
$T(\frak{k})/J,$ then the creation operators are given simply as
right-multiplication by $X.$

I should emphasize that in all three cases the creation and annihilation
operators are unbounded. So these operators are defined only on some (dense)
domain in the corresponding Hilbert space. Although understanding the domains
of these operators can be tricky, I will not concern myself with domain issues here.

\ 

Let us assume now that $K$ (and therefore also $K_{\mathbb{C}}$) is simply
connected. Then between any two of these spaces we have a natural unitary map.
It is almost immediately evident that these unitary maps intertwine the
annihilation operators. For example, the Segal-Bargmann transform $B_{t}$
satisfies
\begin{equation}
B_{t}a_{X}=a_{X}B_{t},\quad X\in\frak{k}.\label{intertwine.1}%
\end{equation}
This is because the Laplacian and also the heat operator is bi-invariant, and
therefore commutes with the left-invariant vector fields $X$. The unitarity of
$B_{t}$ then implies that
\begin{equation}
B_{t}a_{X}^{\ast}=a_{X}^{\ast}B_{t},\quad X\in\frak{k}.\label{intertwine.2}%
\end{equation}
So the Segal-Bargmann transform ``intertwines'' both the creation and the
annihilation operators. It also takes the vacuum to the vacuum, as is easily
verified. Similar statements apply to the Taylor map, and therefore also to
the Hermite expansion map.

It is not hard to see that $B_{t}$ is the unique map (up to constant)
satisfying (\ref{intertwine.1}). Similar statements apply to the Taylor map
and to the Hermite expansion. Indeed this intertwining property was the way in
which Gross originally described the Hermite expansion in \cite[Thm. 2.1]{G2}.
See also Corollary 2.11 and Proposition 7.5 of \cite{G9}.

\begin{problem}
Explain the existence of these unitary intertwining maps.
\end{problem}

This is a vaguely stated problem, which I now explain in more detail. I will
also discuss below one possible solution to this problem. Suppose we have a
Hilbert space $H$ and a collection of ``annihilation operators'' $\left\{
a_{X}\right\}  _{X\in\frak{k}}$ such that the following two properties hold.

\begin{quote}
(A) For all $X,Y\in\frak{k}$, $\left[  a_{X},a_{Y}\right]  =a_{\left[
X,Y\right]  }.$

(B) There is a unique (up to a constant) unit vector $\psi_{0}\in H$ such that
$a_{X}\psi_{0}=0$ for all $X.$
\end{quote}

\noindent Now suppose we have another such space with its own annihilation
operators and vacuum state. Under what conditions does there exist a unitary
map between the two Hilbert spaces that intertwines the vacuum and the
annihilation operators? The preceding discussion says that if $K$ is simply
connected then such unitary intertwining maps do exist between any two of the
three spaces $L^{2}(K,\rho_{t}),$ $\mathcal{H}L^{2}(K_{\mathbb{C}},\mu_{t}),$
and $J_{t}^{0}.$ On the other hand, since (A) and (B) do not impose any
restriction on the inner product on $H,$ one clearly cannot expect a
\textit{unitary} intertwining map to exist in general. Thus it is natural to
look for some additional properties, beyond (A) and (B), that hold in these
three spaces and that could ``explain'' the existence of the unitary
intertwining maps.

It is reasonable to take as our ``target'' Hilbert space the Fock space
$J_{t}^{0},$ since this space does not depend on whether $K$ is simply
connected or not and since it is constructed algebraically using just the
commutation relations (A). So we may ask: Under what conditions on the space
$H$ and the operators $a_{X}$ will there exist a unitary map $U$ from $H$ to
$J_{t}^{0}$ such that $U\psi_{0}=\phi_{0}$ and such that $Ua_{X}=a_{X}U$ for
all $X?$ The goal is to find general conditions (in addition to (A) and (B))
that would guarantee the existence of such a map. If such conditions could be
seen to hold in $L^{2}(K,\rho_{t})$ and $\mathcal{H}L^{2}(K_{\mathbb{C}}%
,\mu_{t})$ then these conditions would ``explain'' why the annihilation
operators in those two spaces are unitarily equivalent to the annihilation
operators in $J_{t}^{0}.$

If there is a unitary map $U$ from $H$ into $J_{t}^{0}$ such that $U\psi
_{0}=\phi_{0}$ and $Ua_{X}=a_{X}U$ then $U$ must satisfy
\begin{equation}
U\left(  a_{X_{1}}^{\ast}\cdots a_{X_{n}}^{\ast}\psi_{0}\right)  =a_{X_{1}%
}^{\ast}\cdots a_{X_{n}}^{\ast}\phi_{0}.\label{u.def}%
\end{equation}
So we have no choice as to what $U$ does on the ``excited states'' $a_{X_{1}%
}^{\ast}\cdots a_{X_{n}}^{\ast}\psi_{0}.$ Assuming a natural independence
condition among the excited states in $H$ we can simply \textit{define} $U$ by
(\ref{u.def}) on the span of the excited states. (This span must be dense in
$H$ if a unitary map $U$ with the desired properties is to exist.) We must
then answer two questions. First, under what conditions on the system
$(H,\left\{  a_{X}\right\}  )$ will the map $U$ defined in (\ref{u.def}) be
isometric? Second, under what conditions will $U$ map onto $J_{t}^{0}$?

In the case $K=\mathbb{R}^{d},$ we do have conditions that will guarantee
affirmative answers to both questions, namely the \textbf{canonical
commutation relations}:
\begin{equation}
\left[  a_{X},a_{Y}^{\ast}\right]  =t\left\langle X,Y\right\rangle
I.\label{ccrs}%
\end{equation}
This relation is in addition to the commutative case of (A), which tells us
that $\left[  a_{X},a_{Y}\right]  =\left[  a_{X}^{\ast},a_{Y}^{\ast}\right]
=0 $. The relation (\ref{ccrs}) can be verified in each of the three Hilbert
spaces $L^{2}(\mathbb{R}^{d},\rho_{t}),$ $\mathcal{H}L^{2}(\mathbb{C}^{d}%
,\mu_{t}),$ and $I_{t}^{0}.$ The \textbf{Stone-von Neumann theorem} tells us
(assuming irreducibility and some domain conditions) that there is a unique
intertwining map between any two Hilbert spaces satisfying the canonical
commutation relations (\ref{ccrs}) and $\left[  a_{X},a_{Y}\right]  =\left[
a_{X}^{\ast},a_{Y}^{\ast}\right]  =0.$ Concretely, one can use the canonical
commutation relations inductively to compute the inner products of different
excited states. One gets a definite answer independent of the particular
representation of the canonical commutation relations, which shows that the
map defined in (\ref{u.def}) is isometric (indeed unitary assuming
irreducibility). (See \cite{H6,RS}.)

So in the $\mathbb{R}^{d}$ case the canonical commutation relations are the
information we need, in addition to (A) and (B), to guarantee that $U$ in
(\ref{u.def}) is unitary. However, for general Lie groups of compact type we
do not have any simple analog of the canonical commutation relations.
\textit{Among }the annihilation operators we have the relation $\left[
a_{X},a_{Y}\right]  =a_{\left[  X,Y\right]  }$ and \textit{among} the creation
operators we have $\left[  a_{X}^{\ast},a_{Y}^{\ast}\right]  =-a_{\left[
X,Y\right]  }^{\ast}.$ But \textit{between} a creation operator and an
annihilation operator things are not so simple. In $L^{2}(K,\rho_{t})$ we can
compute that $\left[  a_{X},a_{Y}^{\ast}\right]  =-\left[  X,Y\right]
-XY(\log\rho_{t}).$ In the $\mathbb{R}^{d}$ case $\left[  X,Y\right]  =0$ and
$XY(\log\rho_{t})$ is a constant, and we recover the canonical commutation
relations. In general, the second derivatives of the logarithm of the heat
kernel are non-constant. Thus $\left[  a_{X},a_{Y}^{\ast}\right]  $ is neither
a multiple of the identity nor any other operator that can be expressed in
terms of the creation and annihilation operators---it is just another
algebraically unrelated operator. Higher commutators, such as $\left[
a_{X},\left[  a_{Y},a_{Z}^{\ast}\right]  \right]  ,$ involve higher
derivatives of the heat kernel, which again are not expressible in terms of
the commutators we already have. There is therefore no simple analog of
(\ref{ccrs}) for general Lie groups of compact type.

There is however, another relation that does hold in general and might be a
partial substitute for the missing canonical commutation relations. Let
$\left\{  X_{k}\right\}  $ be an orthonormal basis for $\frak{k},$ and let
$a_{k}:=a_{X_{k}}$ be the associated annihilation operators. Then I claim
that
\begin{equation}
I=\sum_{n=0}^{\infty}\frac{t^{n}}{n!}\sum_{k_{1},\cdots,k_{n}=1}^{\dim
\frak{k}}a_{k_{1}}^{\ast}\cdots a_{k_{n}}^{\ast}\left|  \psi_{0}\right\rangle
\left\langle \psi_{0}\right|  a_{k_{n}}\cdots a_{k_{1}}.\label{create.id}%
\end{equation}
Here the $a_{k}$'s are the annihilation operators in any one of the three
Hilbert spaces $L^{2}(K,\rho_{t}),$ $\mathcal{H}L^{2}(K_{\mathbb{C}},\mu_{t}),
$ or $J_{t}^{0},$ and $\psi_{0}$ is the vacuum state in that Hilbert space.
The operator $\left|  \psi_{0}\right\rangle \left\langle \psi_{0}\right|  $ is
the orthogonal projection onto the vacuum state. It is not too hard to verify
this relation at least formally in each of the three cases.

My conjecture is that if we have a Hilbert space $H$ satisfying (A) and (B)
above, and also satisfying (\ref{create.id}) plus certain domain conditions,
then there will exist a unique isometric embedding of $H$ \textit{into}
$J_{t}^{0}$ which intertwines the annihilation operators and the vacuum. If
this is correct then one would seek further conditions under which $H$ maps
\textit{onto} $J_{t}^{0}.$ One necessary condition for surjectivity is that
there be no extra relations among the annihilation operators (besides (A)).
That is, we need that the algebra generated by the annihilation operators be
isomorphic to the full universal enveloping algebra. Unfortunately, this
condition is not sufficient to distinguish between $L^{2}(K,\rho_{t})$ with
$K$ simply connected (which maps onto $J_{t}^{0}$) and $L^{2}(K,\rho_{t})$
with $K$ not simply connected (which maps to a proper subspace of $J_{t}^{0}%
$). So some stronger condition is needed.

\section{The $K$-invariant Segal--Bargmann transform}

\subsection{The transform}

Recall that $\mu_{t}$ is the heat kernel (at the identity) on $K_{\mathbb{C}%
}.$ Now consider the function $\nu_{t}$ on $K_{\mathbb{C}}$ given by
\[
\nu_{t}\left(  g\right)  :=\int_{K}\mu_{t}\left(  gx\right)  \,dx.
\]
This function is just the heat kernel (at the identity coset) on the symmetric
space $K_{\mathbb{C}}/K,$ regarded as a $K$-invariant function on
$K_{\mathbb{C}}.$ We also consider the associated measure
\[
d\nu_{t}\left(  g\right)  :=\nu_{t}\left(  g\right)  \,dg.
\]

\begin{theorem}
\label{sb.thm3}For all $t>0$ the map $C_{t}$ given by
\[
C_{t}f=\text{ analytic continuation of }e^{t\Delta_{K}/2}f
\]
is a unitary map of $L^{2}(K,dx)$ onto $\mathcal{H}L^{2}(K_{\mathbb{C}}%
,\nu_{t}).$
\end{theorem}

This result (for the compact case) is Theorem 2 of \cite{H1}. Note that the
map $C_{t}$ is given by precisely the same formula as the transform $B_{t}$;
only the measures on $K$ and $K_{\mathbb{C}}$ have changed.

In the $\mathbb{R}^{d}$ case, the transform $C_{t}$ is nothing but the usual
Segal-Bargmann transform, with trivial differences of normalization. For
example, comparing to Bargmann's transform $A$ we have simply (with $t=1$ as
in \cite{B})
\[
C_{1}f\left(  z\right)  =c\,e^{-z^{2}/4}Af\left(  \tfrac{z}{\sqrt{2}}\right)
,
\]
where $c$ is a constant whose value is unimportant. In turn Bargmann's
transform differs from the finite-dimensional version of Segal's \cite{S3}
essentially just by the ``ground state transformation.'' In the general group
case, $C_{t}$ and $B_{t}$ are ``inequivalent.'' Nevertheless, in light of the
above observations, we may say that in the $\mathbb{R}^{d}$ case both reduce
to the classical Segal-Bargmann transform.

For certain problems the $C_{t}$ version of the transform for $K$ is
preferable to the $B_{t}$ version, mainly because it is invariant in a natural
way under the left and right action of $K.$ This section describes results
that are (except in the $\mathbb{R}^{d}$ case) specific to the $K$-invariant
version of the Segal-Bargmann transform.

\subsection{The connection with geometric quantization}

In this section I summarize the results of a forthcoming paper \cite{H8}. The
main result is that both the Segal-Bargmann space $\mathcal{H}L^{2}%
(K_{\mathbb{C}},\nu_{t})$ and the associated Segal-Bargmann transform $C_{t}$
may be obtained in a natural way by means of geometric quantization. This
generalizes a well-known example from the $\mathbb{R}^{d}$ case \cite[Sect.
9.5]{Wo}.

Geometric quantization gives a method for constructing a ``quantum'' Hilbert
space from a symplectic manifold $\left(  \mathcal{M},\omega\right)  $ (the
``classical phase space''). In order to carry out the quantization
$\mathcal{M}$ must satisfy a certain integrality condition, which is always
satisfied if the symplectic form $\omega$ is exact, as it will be in the case
we will discuss here. The quantization procedure depends on a parameter
$\hbar$ (Planck's constant), which in our case is to be identified with the
parameter $t$ in $C_{t}.$ The quantization also depends on the choice of a
``polarization.'' Roughly a polarization is a choice of $d$ coordinates on the
$2d$-dimensional phase space $\mathcal{M},$ with the idea that the functions
in our quantum Hilbert space will be independent of these $d$ variables. For
example if $\mathcal{M}=\mathbb{R}^{2d},$ then we may take the usual position
and momentum variables $x_{1},\cdots,x_{d},p_{1},\cdots,p_{d}$ and then
consider functions that depend only on $x_{1},\cdots,x_{d}$ and are
independent of $p_{1},\cdots,p_{d}.$ (This is called the vertical
polarization.) In that case our Hilbert space will consist of functions of
$x_{1},\cdots,x_{d}$: the position Hilbert space. Alternatively, one may
consider complex variables $z_{1},\cdots z_{d},\bar{z}_{1},\cdots,\bar{z}_{d}$
and then consider the functions that are ``independent of $\bar{z}_{k}$,''
that is, holomorphic. (This is a complex polarization.) In that case our
Hilbert space is the Segal-Bargmann space.

To be more precise, in geometric quantization the elements of the quantum
Hilbert space are not functions, but rather sections of a certain complex line
bundle with connection. The sections are required to be covariantly constant
in the directions corresponding to the polarization. The Gaussian measure in
the Segal-Bargmann space arises naturally out of the bundle constructions
\cite[Sect. 9.2]{Wo}. See \cite{Wo} and \cite{Ki} for general information on
geometric quantization and polarizations.

If one has two different polarizations on the same symplectic manifold, then
there is a natural ``pairing map'' that maps between the two associated
Hilbert spaces. This map will not be unitary in general, but it is unitary in
certain very special cases. The prime example is $\mathcal{M}=\mathbb{R}^{2d}
$ (with the standard symplectic form). If one uses the ``vertical
polarization'' then the resulting Hilbert space can be identified with
$L^{2}(\mathbb{R}^{d}),$ the ``position Hilbert space.'' If one uses a
``complex polarization'' then the resulting Hilbert space is the
Segal-Bargmann space, with the parameter $t$ now playing the role of Planck's
constant. In this case the pairing map between the two Hilbert spaces is just
the Segal-Bargmann transform. In particular the pairing map is unitary in this
case. (The exact form of the spaces and the transform depends on certain
choices, namely the choice of a trivializing section of the relevant line bundles.)

I now wish to consider $\mathcal{M}=T^{\ast}(K),$ the cotangent bundle of a
connected Lie group of compact type. As a cotangent bundle $T^{\ast}(K)$ has a
canonical symplectic structure. I will describe in the next paragraph how
$T^{\ast}(K)$ can be given a certain complex polarization, obtained by
identifying $T^{\ast}(K)$ with $K_{\mathbb{C}}$. It turns out that the result
of geometric quantization of $T^{\ast}(K)$ (with this complex polarization)
can be identified precisely with the $K$-invariant form of the Segal-Bargmann
space, $\mathcal{H}L^{2}\left(  K_{\mathbb{C}},\nu_{t}\right)  ,$ provided
that one includes the ``half-form correction'' in the quantization. Not only
so, but the pairing map between the complex-polarized Hilbert space and the
vertically-polarized Hilbert space turns out to coincide exactly with the
$K$-invariant Segal-Bargmann transform. In particular, the pairing map is
unitary in this case. These results are remarkable because the Segal-Bargmann
space and Segal-Bargmann transform are defined in terms of heat kernels,
whereas geometric quantization seemingly has nothing to do with heat kernels
or the heat equation.

I will now describe these results in greater detail. The polar decomposition
for $K_{\mathbb{C}}$ gives a diffeomorphism between $K_{\mathbb{C}}$ with the
cotangent bundle $T^{\ast}(K).$ This diffeomorphism is natural from at least
two different points of view, as explained in \cite{H3} and \cite{H7}.
Explicitly, the diffeomorphism is as follows. Identify $T^{\ast}(K)$ with
$K\times\frak{k}^{\ast}$ by means of left-translation, and then with
$K\times\frak{k}$ by means of the inner product on $\frak{k}.$ Then consider
the map $\Phi$ from $K\times\frak{k}$ to $K_{\mathbb{C}}$ given by
\begin{equation}
\Phi\left(  x,Y\right)  =xe^{iY},\quad x\in K,\,Y\in\frak{k}.\label{phi.def}%
\end{equation}
Then $\Phi$ is a diffeomorphism whose inverse is essentially the polar
decomposition for $K_{\mathbb{C}}.$

This diffeomorphism allows us to transport the complex structure of
$K_{\mathbb{C}}$ to $T^{\ast}(K).$ The resulting complex structure on
$T^{\ast}(K)$ fits together with the canonical symplectic structure so as to
make $T^{\ast}(K)$ into a K\"{a}hler manifold. Thus in the language of
geometric quantization we have on $K_{\mathbb{C}}$ a \textbf{K\"{a}hler
polarization}. This K\"{a}hler structure on $T^{\ast}(K)$ has a global
K\"{a}hler potential given by
\begin{equation}
\kappa\left(  x,Y\right)  =\left|  Y\right|  ^{2}.\label{kahler.pot}%
\end{equation}
This K\"{a}hler potential satisfies $\operatorname{Im}\bar{\partial}%
\kappa=\theta,$ where $\theta$ is the canonical 1-form on $T^{\ast}(K),$ given
by $\theta=\Sigma p_{k}dq_{k}$ in the usual sort of local coordinates. This
complex structure and the associated K\"{a}hler potential come from the work
of Guillemin and Stenzel and of Lempert and Sz\"{o}ke. (See \cite[Sect.
5]{GStenz} and \cite[Cor. 5.5]{LS}.)

We now follow the prescription of geometric quantization. First we construct a
\textbf{prequantum line bundle} $L$, which is a Hermitian line bundle with
connection whose curvature is equal to $\omega/\hbar,$ where $\omega$ is the
canonical symplectic form on $T^{\ast}(K).$ In our case $\omega=d\theta$ is
exact and so we may choose our line bundle $L$ to be topologically trivial; in
fact we just take $L=T^{\ast}(K)\times\mathbb{C},$ with covariant derivative
given by $\nabla_{X}=X+\frac{1}{i\hbar}\theta\left(  X\right)  $ (see
\cite[Eq. (8.4.4)]{Wo}).

Next we use the K\"{a}hler polarization on $T^{\ast}(K)$ to define the notion
of \textbf{polarized section} of $L.$ These are the sections that are
covariantly constant in the $\bar{z}$-directions. Since we are using a
K\"{a}hler polarization, $L$ obtains the structure of a holomorphic line
bundle, and the polarized sections are precisely the holomorphic sections.
Since our line bundle is trivial, the sections are just functions. The
holomorphic sections, however, are not the same as holomorphic functions.
Rather, the holomorphic sections are functions of the form $s=\exp\left(
-\kappa/2\hbar\right)  F,$ where $F$ is a holomorphic function and $\kappa$ is
the K\"{a}hler potential (\ref{kahler.pot}).

Finally, we consider the \textbf{half-form correction}. This means that we
consider the canonical bundle for $T^{\ast}(K)$ (relative to the K\"{a}hler
polarization), namely, the holomorphic line bundle whose holomorphic sections
are holomorphic $d$-forms on $T^{\ast}(K).$ (Here $d=\dim K=\frac{1}{2}\dim
T^{\ast}(K).$) This bundle has a natural Hermitian structure given by $\left|
\alpha\right|  ^{2}=\left(  i^{d}\alpha\wedge\bar{\alpha}\right)  /\omega
^{d}.$ In our case, the canonical bundle is holomorphically trivial, since
thinking of $T^{\ast}(K)$ as $K_{\mathbb{C}}$ we have a nowhere-vanishing
holomorphic $d$-form given by $\alpha=\eta_{1}\wedge\cdots\wedge\eta_{d},$
where the $\eta_{k}$'s are left-$K_{\mathbb{C}}$-invariant holomorphic
1-forms. We then consider a bundle $\delta,$ which is supposed to be a
\textbf{square root of the canonical bundle}. In this case we may take
$\delta$ to be a holomorphically trivial bundle whose square is isomorphic to
the canonical bundle. Then $\delta$ inherits a Hermitian structure from the
canonical bundle, defined by requiring that if $s$ is a section of $\delta,$
then $\left|  s\right|  =\left|  s\otimes s\right|  ^{1/2}.$

So then we consider the holomorphic line bundle $L\otimes\delta,$ which has a
natural Hermitian structure, since both $L$ and $\delta$ do. The
\textbf{quantum Hilbert space} is then the space of holomorphic sections $s$
of $L\otimes\delta$ that are square-integrable with respect to the natural
volume measure $\omega^{d}$ on $T^{\ast}(K).$ For such a section $s$ we define
$\left\|  s\right\|  $ by
\[
\left\|  s\right\|  ^{2}=\int_{T^{\ast}(K)}\left|  s\right|  ^{2}\,\omega
^{d}.
\]
In our case, both $L$ and $\delta$ are holomorphically trivial, with
trivializing sections $\exp\left(  -\kappa/2\hbar\right)  $ for $L$ and
$\sqrt{\alpha}$ for $\delta.$ Thus every holomorphic section is of the form
$s=\exp\left(  -\kappa/2\hbar\right)  \sqrt{\alpha}F,$ where $F$ is a
holomorphic function. We may then identify the section $s$ with the function
$F,$ with the norm as follows
\[
\left\|  F\right\|  ^{2}=\int_{T^{\ast}(K)}\left|  s\right|  ^{2}\,\omega
^{d}=\int_{T^{\ast}(K)}\left|  F\right|  ^{2}e^{-\kappa/\hbar}\left|
\frac{i^{d}\alpha\wedge\bar{\alpha}}{\omega^{d}}\right|  ^{1/2}\omega^{d}.
\]

\begin{proposition}
For each $\hbar>0$ there exists a constant $c_{\hbar}$ so that under the
identification of $T^{\ast}(K)$ with $K_{\mathbb{C}}$ as above we have
\[
e^{-\kappa/\hbar}\left|  \frac{i^{d}\alpha\wedge\bar{\alpha}}{\omega^{d}%
}\right|  ^{1/2}\omega^{d}=c_{\hbar}\nu_{\hbar}\left(  g\right)  \,dg.
\]
\end{proposition}

This is to be compared with Section 7 of \cite{H4} in which a similar
calculation was performed, but without inclusion of the half-form correction.
I am grateful to Dan Freed who suggested to me that the discrepancy in
\cite{H4} might be accounted for by the half-form correction. Recall that in
this subsection we are identifying the parameter $t$ in $C_{t} $ with $\hbar.$

To summarize, the $K$-invariant Segal-Bargmann space $\mathcal{H}L^{2}\left(
K_{\mathbb{C}},\nu_{\hbar}\left(  g\right)  dg\right)  $ (up to an irrelevant
overall constant) may be obtained by geometric quantization using a complex
polarization and the half-form correction.

It may also be shown that the $K$-invariant form of the Segal-Bargmann
transform coincides (up to an overall constant) with the pairing map of
geometric quantization. In particular, the pairing map is (a multiple of) a
unitary map in this case. More precisely, computing a formula for the pairing
map similar to Equation (9.5.3) of \cite{Wo}, one sees that the pairing map
coincides with the \textit{inverse} Segal-Bargmann transform as given in
\cite{H2}. Details of these results will appear in the forthcoming paper
\cite{H8}.

\begin{problem}
Explain why the pairing map of geometric quantization is unitary in this case.
\end{problem}

This is the only case I know of, besides examples on $\mathbb{R}^{2d},$ in
which a pairing map is unitary. In examples on $\mathbb{R}^{2d}$ the Stone-von
Neumann theorem can explain why various pairing maps are unitary. In cases
where the Stone-von Neumann theorem is inapplicable, there seems to be no
reason to expect unitarity. The only non-trivial example besides the one
discussed in this section where unitarity has been examined is one involving
spheres investigated by J. Rawnsley \cite{R}, who found that the pairing map
was not unitary. Thus it surprising that the pairing map for $T^{\ast}(K)$ is
unitary, and it seems that there must be some ``explanation'' for this
phenomenon. (See the comments in Section 4.4 of \cite{Ki}.)

\subsection{Phase bounds}

Suppose $f\in L^{2}(K,dx)$ has norm one. Then since the $K$-invariant
Segal-Bargmann transform $C_{t}$ is unitary, $C_{t}f$ will be an element of
norm one in $\mathcal{H}L^{2}(K_{\mathbb{C}},\nu_{t})$. This means by
definition that
\begin{equation}
\int_{K_{\mathbb{C}}}\left|  C_{t}f\left(  g\right)  \right|  ^{2}\nu
_{t}\left(  g\right)  \,dg=1.\label{density.1}%
\end{equation}
Now in the last section we saw that the complex group $K_{\mathbb{C}}$ may be
identified with the cotangent bundle $T^{\ast}(K).$ Using this identification
of $K_{\mathbb{C}}$ with $T^{\ast}(K)$ (given in (\ref{phi.def})) we may
re-write (\ref{density.1}) in terms of the canonical ``phase volume measure''
$\omega^{d}$ on $T^{\ast}(K).$ So (\ref{density.1}) becomes
\begin{equation}
\int_{K_{\mathbb{C}}}\left|  C_{t}f\left(  g\right)  \right|  ^{2}\nu
_{t}\left(  g\right)  \alpha\left(  g\right)  \,\omega^{d}=1,\label{density.2}%
\end{equation}
where $\alpha$ is the density (under our identification) of $\omega^{d}$ with
respect to $dg.$ It then seems natural to interpret the quantity
\begin{equation}
\left|  C_{t}f\left(  g\right)  \right|  ^{2}\nu_{t}\left(  g\right)
\alpha\left(  g\right) \label{density.3}%
\end{equation}
as the \textbf{phase space probability density} associated to the state $f.$
Certainly this quantity is a probability density on phase space, that is, a
positive function that integrates to one with respect to the natural phase
space volume measure. Experience from the $\mathbb{R}^{d}$ case suggest that
this is in some sense the most natural phase space density that can be
associated to the state $f.$ See \cite{H5,H7} for further discussion. In the
$\mathbb{R}^{d}$ case, the density in (\ref{density.3}) is the ``Husimi
function'' of the quantum state $f.$

If we are to interpret the quantity (\ref{density.3}) as a phase space
probability density, then we must take account of the \textbf{uncertainty
principle}, which says roughly that states cannot be too concentrated in phase
space. (The uncertainty principle is a ``meta-theorem,'' which has many
different precise formulations in different settings. See \cite{FS}.) If
$C_{t}f$ were an arbitrary unit vector in $L^{2}(K_{\mathbb{C}},\nu_{t})$ we
would have trouble with the uncertainty principle, because in that case
(\ref{density.3}) could be concentrated in an arbitrarily small region of the
phase space. Fortunately, $C_{t}f$ is in the holomorphic subspace; this
imposes very precise limits on the concentration of the density
(\ref{density.3}), as given in the following theorem.

\begin{theorem}
\label{phase.thm}For any Lie group $K$ of compact type, there exist constants
$a_{t}$ such that for all $f\in L^{2}(K,dx)$ with norm one we have
\[
\left|  C_{t}f\left(  g\right)  \right|  ^{2}\nu_{t}\left(  g\right)
\alpha\left(  g\right)  \leq a_{t}\left(  2\pi t\right)  ^{-d}
\]
for all points $g\in K_{\mathbb{C}}.$ Here $d=\dim K$. The optimal constants
$a_{t}$ tend to one exponentially fast as $t\rightarrow0.$ If $K=\mathbb{R}%
^{d}$ then we may take $a_{t}=1.$
\end{theorem}

This is Theorem 1 of \cite{H3}. The proof of this result depends on the
extremely good information that is available about the heat kernels on $K$ and
on $K_{\mathbb{C}}/K.$ The theorem says that the phase space probability
density cannot be too big at any one point; since the density integrates to
one, this means that it must be fairly spread out in phase space, in agreement
with the uncertainty principle.

The bounds on the probability density are equivalent to bounds on holomorphic
functions in $\mathcal{H}L^{2}(K_{\mathbb{C}},\nu_{t})$ (see Theorem 2 of
\cite{H3}). These bounds are exponentially better than the Driver-Gross bounds
in Theorem \ref{bounds.thm}. On the other hand, the Driver-Gross bounds hold
in much greater generality and have dimension-independent constants.

\begin{problem}
Explain why the constant in Theorem \ref{phase.thm} tends to one
\textbf{exponentially fast}.
\end{problem}

One way of interpreting Theorem \ref{phase.thm} is as saying that the
normalized reproducing kernel, on the diagonal, is asymptotic \textit{to all
orders} to $\left(  2\pi t\right)  ^{-d}.$ It is perhaps reasonable on general
groups (see for example the sort of results in \cite{E}) to expect the
reproducing kernel on the diagonal to be asymptotic to \textit{leading order}
to $\left(  2\pi t\right)  ^{-d}.$ Since in this case it is actually
asymptotic to all orders, one may look for some special geometric property of
this setting that could account for this.

\subsection{Compact symmetric spaces}

A compact symmetric space is a manifold of the form $K/H$ where $K$ is a
compact Lie group and $H$ is a special sort of closed subgroup, namely, the
fixed-point subgroup of an involution of $K.$ Examples include the spheres
$S^{n}$ and complex projective spaces $\mathbb{CP}^{n}.$ M. Stenzel has
generalized some of the results of \cite{H1,H3} to this setting. One can get a
Segal-Bargmann transform for $K/H$ simply by applying the Segal-Bargmann
transform for $K$ and restricting to $H$-invariant functions, as in
\cite[Sect. 11]{H1}. It is easily seen that this is the same as applying the
heat operator for the symmetric space $K/H$ and then analytically continuing
to $K_{\mathbb{C}}/H_{\mathbb{C}}.$ The range Hilbert space will then be a
Hilbert space of holomorphic functions on $K_{\mathbb{C}}/H_{\mathbb{C}}.$ The
problem with this approach is that it does not give a nice description of the
relevant measure on $K_{\mathbb{C}}/H_{\mathbb{C}}.$ If we consider the
$K$-invariant form $C_{t}$ of the transform, then the measure which arises on
$K_{\mathbb{C}}/H_{\mathbb{C}}$ will be the push-forward under the quotient
map of the measure $\nu_{t}$ on $K_{\mathbb{C}}.$ Now, the measure $\nu_{t}$
is just the heat kernel measure for the non-compact symmetric space
$K_{\mathbb{C}}/K,$ viewed as a $K$-invariant measure on $K_{\mathbb{C}}.$
However, it is not clear \textit{a priori} whether the push-forward of
$\nu_{t}$ under the quotient map is some sort of heat kernel measure.

Fortunately, M. Stenzel has solved this problem. He shows in \cite[Thm. 3]{St}
that the Segal-Bargmann transform for $K/H$ is a unitary map onto the Hilbert
space $\mathcal{H}L^{2}(K_{\mathbb{C}}/H_{\mathbb{C}},\sigma_{t}).$ Here
$\sigma_{t}$ is the heat kernel measure for the \textbf{dual non-compact
symmetric space} to $K/H.$ More precisely, Stenzel shows that $K_{\mathbb{C}%
}/H_{\mathbb{C}}$ is identifiable with $T^{\ast}(K/H).$ He then shows that
each of the fibers of $T^{\ast}(K/H)$ may be identified with a certain
non-compact symmetric space $G/H$, which is dual to $K/H$ in the sense of
\cite[Chap. V.2]{He}. (Here $G$ is a certain subgroup of $K_{\mathbb{C}}.$)
The measure $\sigma_{t}$ is then obtained by integrating over each fiber with
respect to the heat kernel measure on $G/H,$ and then integrating over $K/H$
with respect to the volume measure. For example, suppose that $K=SO(n+1)$ and
$H=SO(n).$ Then $K/H=S^{n}.$ In this case the non-compact symmetric space dual
to $S^{n}$ is $n$-dimensional hyperbolic space $\mathbb{H}^{n}$. Stenzel shows
how each fiber in $T^{\ast}(S^{n})$ can be identified with $\mathbb{H}^{n}.$
The measure $\sigma_{t}$ is then essentially the product of the heat kernel
measure on $\mathbb{H}^{n}$ with the volume measure on $S^{n}.$

Stenzel's paper also extends the inversion formula of \cite{H2} to the setting
of compact symmetric spaces \cite[Thms. 1 and 2]{St}. Indeed Stenzel first
proves the inversion formula and then deduces the unitarity of the
Segal-Bargmann from this. In the case of a 2-sphere results related to those
of Stenzel have been obtained independently and from a very different point of
view by K. Kowalski and J. Rebieli\'{n}ski \cite{KR}.

Now, Stenzel does not prove that the measure $\sigma_{t}$ is the same as the
push-forward from $K_{\mathbb{C}}$ of $\nu_{t}.$ (He works directly at the
level of the symmetric space and does not reduce things to the group case.)
Nevertheless, $\sigma_{t}$ \textit{is} the same as the push-forward from
$K_{\mathbb{C}}$ of $\nu_{t}.$ This is proved by F. Zhu \cite{Z}, using
methods of M. Flensted-Jensen. The idea of \cite{F-J} is that one can relate
differential operators on $G/K$ to differential operators on $K_{\mathbb{C}%
}/K;$ in particular it allows us to relate the heat equation on $G/K$ to the
heat equation on $K_{\mathbb{C}}/K.$ This is what we need to relate the
push-forward of $\nu_{t}$ to $\sigma_{t}.$

\begin{problem}
Prove phase space bounds similar to those of Section 3.3 in which the group
$K$ of compact type is replaced by a compact symmetric space.
\end{problem}

It seems reasonable to expect that similar results (perhaps slightly weaker)
will hold. However, the information about heat kernels that is needed is much
harder to obtain in the general case than in the compact group case. That is,
the dual pair of symmetric spaces $K$ and $K_{\mathbb{C}}/K$ is very special
among all dual pairs. As a result there are special formulas for the heat
kernel on these two spaces (see \cite{H3}) that are not available for general
pairs. Thus one has to work harder in the general case.

\begin{problem}
Determine whether or not the constructions of geometric quantization coincide
with the Segal-Bargmann space and Segal-Bargmann transform for $K/H. $ If not,
explain why geometric quantization gives the same result as the ``heat
kernel'' approach in the case of compact groups but not in the case of compact
symmetric spaces.
\end{problem}

I believe that in fact the geometric quantization constructions will
\textit{not} agree with those of Stenzel for general compact symmetric spaces.
Again, the group case is special. It would be very nice of have a good
geometrical explanation of why the two constructions agree in some cases but
not others.

\begin{problem}
Find a unitary Segal-Bargmann transform starting on a \textbf{non-compact}
symmetric space.
\end{problem}

That is, consider a system whose ``position Hilbert space'' is $L^{2}(X)$,
where $X$ is a non-compact symmetric space such as hyperbolic space. If one
simply tries to imitate the constructions from the compact case, one runs into
very serious trouble; for example, the heat kernel simply does not have an
analytic continuation of the expected sort. Although it is possible that there
simply is no theorem here to be proved, I feel that this case should be
understood better. Possibly we need a different way to think about the problem.

\section{Connections with the theory for infinite-dimensional linear spaces}

\subsection{Embedding the isomorphisms}

The results described above for Lie groups had their beginning in the Hermite
expansion for $K,$ as described in Section 2.3. Theorem \ref{hermite.thm2} was
first proved (in a somewhat different form) by L. Gross in \cite{G2}. However,
the Hermite expansion for $K$ was not the main purpose of \cite{G2}. Rather,
the Hermite expansion was a fortuitous discovery along the path to a result in
stochastic analysis that I discuss below. In \cite{G2} Gross thinks of
$L^{2}(K,\rho_{t})$ as a certain special subspace of the space of functions on
an \textit{infinite-dimensional }linear space. The \textit{generalized}
Hermite expansion for $K$ is actually then a special case of the
\textit{ordinary} Hermite expansion for an infinite-dimensional linear space
(the $d\rightarrow\infty$ limit of the results in the introduction). Papers of
Gross and Malliavin \cite{GM}, Hall and Sengupta \cite{HS}, Wren \cite{Wr},
and Driver and Hall \cite{DH} have helped to understand better how the
isomorphisms associated to a compact group relate to the corresponding
isomorphisms for an infinite-dimensional linear space. For additional
exposition of various of these relations see \cite{G9} and also
\cite{G3,G4,H7,L2}.

What happens is that each of the three Hilbert spaces associated to a compact
Lie group can be embedded isometrically as a (very small) subspace of the
corresponding Hilbert space for an infinite-dimensional linear space. Under
this embedding the isomorphisms for the compact group match up with
corresponding isomorphisms for the linear case as in Theorems \ref{gm.thm} and
\ref{chaos} below. Thus each of the isomorphisms for the group case can be
understood as a special case of the corresponding isomorphism for an
infinite-dimensional linear space. It is helpful to have these two
complementary ways of viewing the isomorphisms for $K,$ either directly at the
finite-dimensional level in a way that is merely analogous to the linear case,
or actually as a special case of the isomorphisms for an infinite-dimensional
linear space. For example, the pointwise bounds in Theorem \ref{bounds.thm}
can be understood as a consequence of the embedding and the well-known bounds
from the linear case.

Suppose as usual that $K$ is a connected Lie group of compact type. Let
$\frak{k}$ be the Lie algebra of $K,$ with a fixed Ad-$K$-invariant inner
product. Then consider the real Hilbert space
\[
H_{\mathbb{R}}:=L^{2}(\left[  0,1\right]  ;\frak{k)},
\]
where the inner product on $H_{\mathbb{R}}$ is computed using the inner
product on $\frak{k}.$ Consider also the associated complex Hilbert space
\[
H_{\mathbb{C}}:=L^{2}(\left[  0,1\right]  ;\frak{k}_{\mathbb{C}}).
\]
Then $H_{\mathbb{C}}=H_{\mathbb{R}}+iH_{\mathbb{R}}.$ We wish to think of
$H_{\mathbb{R}} $ as playing the role of $\mathbb{R}^{d}$ and $H_{\mathbb{C}}$
the role of $\mathbb{C}^{d},$ where now $d=\infty.$

We may attempt to construct a Gaussian measure, say $P_{t},$ on $H_{\mathbb{R}%
}, $ which should be given heuristically by the formula
\[
dP_{t}\left(  A\right)  =b_{t}e^{-\left\|  A\right\|  ^{2}/2t}\mathcal{D}A.
\]
Here $\mathcal{D}A$ is the non-existent Lebesgue measure on $H_{\mathbb{R}},$
and $b_{t}$ is supposed to be a normalization constant. It is well known
\cite{G1} that $P_{t}$ may be given a rigorous meaning as a probability
measure on a suitable ``extension'' of $H_{\mathbb{R}},$ denoted $\bar
{H}_{\mathbb{R}}.$ For example, $\bar{H}_{\mathbb{R}}$ may be taken to be the
space of $\frak{k}$-valued distributions on $\left[  0,1\right]  .$ The
original space $H_{\mathbb{R}}$ is a set of $P_{t}$-measure zero inside
$\bar{H}_{\mathbb{R}}.$ The measure $P_{t}$ describes $\frak{k}$-valued white
noise (scaled by a factor of $\sqrt{t}$), or equivalently, the derivative of
scaled $\frak{k}$-valued Brownian motion. Similarly we may construct a
Gaussian measure $M_{t} $ on an extension $\bar{H}_{\mathbb{C}}$ of
$H_{\mathbb{C}},$ where heuristically
\[
dM_{t}\left(  Z\right)  =c_{t}e^{-\left\|  Z\right\|  ^{2}/t}\mathcal{D}Z.
\]

Then we have the position Hilbert space
\[
L^{2}(\bar{H}_{\mathbb{R}},P_{t})
\]
and the Segal-Bargmann space
\[
\mathcal{H}L^{2}(\bar{H}_{\mathbb{C}},M_{t}).
\]
There are technical subtleties in the definition of the Segal-Bargmann space,
which I will not discuss here. (See for example \cite{Su}.) Finally we have
the dual of the symmetric algebra over $H_{\mathbb{R}},$ denoted
\[
I_{t}^{0}(H_{\mathbb{R}}).
\]
By a straightforward limit $d\rightarrow\infty$ we may extend the results for
$\mathbb{R}^{d}$ to this setting, and so obtain unitary maps
\begin{align*}
B_{t}  & :L^{2}(\bar{H}_{\mathbb{R}},P_{t})\rightarrow\,\mathcal{H}L^{2}%
(\bar{H}_{\mathbb{C}},M_{t})\\
\text{Taylor}  & :\mathcal{H}L^{2}(\bar{H}_{\mathbb{C}},M_{t})\rightarrow
I_{t}^{0}(H_{\mathbb{R}}).
\end{align*}
The composition of the Segal-Bargmann transform and the Taylor map is the
infinite-dimensional version of the Hermite expansion, which in this setting
takes the form of an expansion into multiple stochastic integrals, as
discussed below.

Now consider the \textbf{holonomy map}
\[
h:\bar{H}_{\mathbb{R}}\rightarrow K
\]
given by
\begin{equation}
h\left(  A\right)  =\lim_{n\rightarrow\infty}e^{\int_{0}^{1/n}A(\tau)d\tau
}e^{\int_{1/n}^{2/n}A(\tau)d\tau}\cdots e^{\int_{(n-1)/n}^{1}A(\tau)d\tau
}.\label{holonomy.form}%
\end{equation}
It is possible to show that $h\left(  A\right)  $ is defined for $P_{t}%
$-almost every $A\in\bar{H}_{\mathbb{R}}.$ More conventionally, $h\left(
A\right)  $ may be defined as the solution at time one of a certain stochastic
differential equation, called the It\^{o} map. (See (\ref{holonomy.ito})
below.) Similarly we have
\[
h_{\mathbb{C}}:\bar{H}_{\mathbb{C}}\rightarrow K_{\mathbb{C}}
\]
given by
\[
h\left(  Z\right)  =\lim_{n\rightarrow\infty}e^{\int_{0}^{1/n}Z(\tau)d\tau
}e^{\int_{1/n}^{2/n}Z(\tau)d\tau}\cdots e^{\int_{(n-1)/n}^{1}Z(\tau)d\tau}.
\]
Formally $h_{\mathbb{C}}$ is just the analytic continuation of $h$ from
$\bar{H}_{\mathbb{R}}$ to $\bar{H}_{\mathbb{C}}.$ I will explain below a bit
more the origin of this map.

\begin{proposition}
For all $t>0$ the map
\[
\phi\rightarrow\phi\circ h
\]
is an isometric embedding of $L^{2}(K,\rho_{t})$ into $L^{2}(\bar
{H}_{\mathbb{R}},P_{t}).$ For all $t>0$ the map
\[
\Phi\rightarrow\Phi\circ h_{\mathbb{C}}
\]
is an isometric embedding of $\mathcal{H}L^{2}(K_{\mathbb{C}},\mu_{t})$ into
$\mathcal{H}L^{2}(\bar{H}_{\mathbb{C}},M_{t}).$
\end{proposition}

So we are embedding $L^{2}(K,\rho_{t})$ into $L^{2}(\bar{H}_{\mathbb{R}}%
,P_{t}) $ as the space of functions of the form
\begin{equation}
f\left(  A\right)  =\phi\left(  h\left(  A\right)  \right)  ,\label{f.form}%
\end{equation}
where $\phi$ is a function on $K.$ The map $\phi\rightarrow\phi\circ h$ is
isometric because the push-forward of the Gaussian measure $P_{t}$ under $h$
is precisely the heat kernel measure $\rho_{t}.$ (In probabilistic language,
the distribution of $h\left(  A\right)  $ with respect to $P_{t}$ is $\rho
_{t}.$) Similarly we are embedding $\mathcal{H}L^{2}(K_{\mathbb{C}},\mu_{t})$
into $\mathcal{H}L^{2}(\bar{H}_{\mathbb{C}},M_{t})$ as the space of functions
of the form $F\left(  Z\right)  =\Phi\left(  h_{\mathbb{C}}\left(  Z\right)
\right)  ,$ which is isometric because the push-forward of $M_{t}$ under
$h_{\mathbb{C}}$ is $\mu_{t}.$ A technical issue that must be resolved is to
show that for $\Phi\in\mathcal{H}L^{2}\left(  K_{\mathbb{C}},\mu_{t}\right)
,$ $F=\Phi\circ h_{\mathbb{C}}$ is holomorphic in the appropriate sense on
$\bar{H}_{\mathbb{C}}.$ (See \cite[Sect. 2.5]{HS}.)

\begin{theorem}
\label{gm.thm}Suppose that $f\in L^{2}(\bar{H}_{\mathbb{R}},P_{t})$ is of the
form $f\left(  A\right)  =\phi\left(  h\left(  A\right)  \right)  ,$ where
$\phi$ is a function on $K.$ Then
\[
B_{t}f\left(  Z\right)  =\Phi\left(  h_{\mathbb{C}}\left(  Z\right)  \right)
\]
where $\Phi$ is the analytic continuation to $K_{\mathbb{C}}$ of
$e^{t\Delta_{K}/2}\phi.$ In other words, on functions of the form $\phi\circ
h,$ the ordinary Segal-Bargmann transform for the infinite-dimensional linear
space $\bar{H}_{\mathbb{R}}$ reduces to the generalized Segal-Bargmann
transform for the group $K.$
\end{theorem}

This result (in a slightly different form) is due to Gross and Malliavin
\cite[Cor. 7.12]{GM}. See also \cite[Sect. 2.5]{HS}.

Meanwhile, let us consider the infinite-dimensional version of the Hermite
expansion, for the space $L^{2}(\bar{H}_{\mathbb{R}},P_{t}).$ This takes the
form of an expansion into multiple stochastic integrals, the \textbf{Wiener
chaos expansion}.

\begin{theorem}
\label{chaos}Suppose that $f$ is a function on $\bar{H}_{\mathbb{R}}$ of the
form $f\left(  A\right)  $ $=\phi\left(  h\left(  A\right)  \right)  .$ Let
$\xi\in J_{t}^{0}(\frak{k})$ be the Hermite expansion of $\phi$ in the sense
of Section 2.4. Then the Hermite expansion of $f$ is given by
\begin{equation}
f\left(  A\right)  =\sum_{n=0}^{\infty}\int_{0}^{1}\int_{0}^{\tau_{n}}%
\cdots\int_{0}^{\tau_{2}}\xi_{n}\left(  da_{\tau_{1}}\otimes\cdots\otimes
da_{\tau_{n}}\right)  .\label{f.expand}%
\end{equation}
Here $a_{\tau}:=\int_{0}^{\tau}A_{\sigma}\,d\sigma$ is the Brownian motion
associated to the white noise $A,$ and the integrals are It\^{o} stochastic integrals.
\end{theorem}

The situation is thus similar to that for the Segal-Bargmann transform. When
applied to functions of the form $\phi\circ f,$ the Hermite expansion for the
linear space $\bar{H}_{\mathbb{R}}$ reduces to the generalized Hermite
expansion for the compact group $K.$ This result is a consequence of Theorems
2.4 and 2.5 of \cite{G2}, together with the explicit formula for the Hermite
expansion as given in \cite[Prop. 2.4]{Hi1}. The result is obtained directly
in Lemma 5.7 of \cite{DH}. One may also compute the Taylor expansion of
functions in $\mathcal{H}L^{2}(\bar{H}_{\mathbb{C}},M_{t})$ of the form
$F\left(  Z\right)  =\Phi\left(  h_{\mathbb{C}}\left(  Z\right)  \right)  ,$
with similar results. See \cite[Lem. 5.8]{DH}.

Note that in the chaos expansion of a general function in $L^{2}(\bar
{H}_{\mathbb{R}},P_{t})$ the integrands $\xi_{n}$ would be functions of the
variables $\tau_{1},\cdots,\tau_{n}.$ Here the integrands are just constants,
and the elements $\xi_{n}$ must fit together to define an element of
$T(\frak{k})^{\ast}$ that annihilates the ideal $J.$

Let me explain briefly the background to these results. Let $W(K)$ denote the
\textbf{continuous path group}
\[
W(K)=\left\{  \left.  \text{maps }x:\left[  0,1\right]  \rightarrow K\right|
x_{0}=e,\text{ }x\text{ is continuous}\right\}  .
\]
There is a natural probability measure $w_{t}$ on $W(K),$ the \textbf{Wiener
measure}, which describes Brownian motion in $K.$ (The parameter $t$ is
\textit{not} the time variable for the Brownian motion, but rather a scaling
factor that determines the diffusion rate of the Brownian motion.) Now let
$\mathcal{L}\left(  K\right)  $ denote the \textbf{finite-energy loop group}:
\[
\mathcal{L}\left(  K\right)  =\left\{  \left.  \text{maps }l:\left[
0,1\right]  \rightarrow K\right|  l_{0}=l_{1}=e,\text{ }l\text{ has one
derivative in }L^{2}\right\}  .
\]
There is a natural right action of $\mathcal{L}\left(  K\right)  $ on $W(K),$
given by $x\rightarrow xl,$ with $x\in W(K),$ $l\in\mathcal{L}\left(
K\right)  .$ This action leaves the Wiener measure \textbf{quasi-invariant};
that is, $dw_{t}\left(  xl\right)  $ is absolutely continuous with respect to
$dw_{t}\left(  x\right)  .$

Given the Wiener measure $w_{t}$, we may consider the Hilbert space
$L^{2}\left(  W(K),w_{t}\right)  .$ We may then consider the functions $f$ in
$L^{2}\left(  W(K),w_{t}\right)  $ that are \textbf{loop-invariant}, that is,
such that for all $l\in\mathcal{L}\left(  K\right)  $ we have
\[
f\left(  xl\right)  =f\left(  x\right)
\]
for $w_{t}$-almost every $x\in W(K).$ One obvious class of loop-invariant
functions is the class of \textbf{endpoint functions}, namely those of the
form
\[
f\left(  x\right)  =\phi\left(  x_{1}\right)  ,
\]
where $\phi$ is a function on $K.$ (Note that by the definition of
$\mathcal{L}\left(  K\right)  ,$ $\left(  xl\right)  _{1}=x_{1}l_{1}=x_{1}.$)

\begin{theorem}
\label{loop.thm}Every loop-invariant function in $L^{2}(W(K),w_{t})$ is an
endpoint function.
\end{theorem}

This result is a consequence of Theorem 2.5 of \cite{G2}. The difficulty in
proving this result is that we consider continuous paths but only
finite-energy loops. The restriction to finite-energy loops is necessary to
have quasi-invariance. Without quasi-invariance the notion of loop-invariance
does not make sense, since the elements of $L^{2}(W(K),w_{t}) $ are not
actually functions but rather \textit{equivalence classes} of functions that
are equal $w_{t}$-almost everywhere.

To prove this theorem Gross linearizes it, transferring the problem from the
path group $W(K)$ to the linear space $\bar{H}_{\mathbb{R}}.$ The transfer is
accomplished by means of the \textbf{It\^{o} map}, which is the
measure-preserving map
\[
\theta:\left(  \bar{H}_{\mathbb{R}},P_{t}\right)  \rightarrow\left(
W(K),w_{t}\right)
\]
obtained by solving the Stratonovich stochastic differential equation
\[
dx_{\tau}=x_{\tau}\circ da_{\tau},\quad x_{0}=e,
\]
where as above $a_{\tau}:=\int_{0}^{\tau}A_{\sigma}\,d\sigma$ is the Brownian
motion associated to the white noise $A.$ Once the equation is solved we set
$\theta\left(  A\right)  =x.$ The holonomy map is nothing but the It\^{o} map
at time one:
\begin{equation}
h\left(  A\right)  =\theta(A)_{1},\quad A\in\bar{H}_{\mathbb{R}}%
.\label{holonomy.ito}%
\end{equation}

The It\^{o} map induces a unitary map of $L^{2}(W(K),w_{t})$ onto $L^{2}%
(\bar{H}_{\mathbb{R}},P_{t}).$ Under this map, the endpoint functions in
$L^{2}(W(K),w_{t})$ go to functions of the holonomy in $L^{2}(\bar
{H}_{\mathbb{R}},P_{t}).$ Furthermore, one may use the It\^{o} map to transfer
the loop-group action on $W(K)$ to $\bar{H}_{\mathbb{R}}.$ The resulting
action is given by
\begin{equation}
\left(  l\cdot A\right)  _{\tau}=l_{\tau}A_{\tau}l_{\tau}^{-1}-\frac{dl}%
{d\tau}l_{\tau}^{-1},\quad l\in\mathcal{L}\left(  K\right)  ,\,A\in\bar
{H}_{\mathbb{R}}.\label{loop.act}%
\end{equation}
So Theorem \ref{loop.thm} is equivalent to the statement that every function
on $\bar{H}_{\mathbb{R}}$ that is invariant under the action (\ref{loop.act})
is of the form $f\left(  A\right)  =\phi\left(  h\left(  A\right)  \right)  ,$
for some function $\phi$ on $K.$ To prove this last statement, Gross proves
that if $f$ is loop-invariant on $\bar{H}_{\mathbb{R}}$ then the chaos
expansion of $f$ must be of the form (\ref{f.expand}), for some $\xi\in
J_{t}^{0}$ \cite[Thm. 5.1]{G2}. Then to construct the function $\phi$ on $K, $
Gross essentially inverts the Hermite expansion for $K$ \cite[Sect. 8]{G2}.

Of course, prior to \cite{G2}, the Hermite expansion for $K$ was not
known---it was discovered by Gross as a consequence of the investigations
described in this section. For a simplified version of Gross's proof of
Theorem \ref{loop.thm}, see Part II of \cite{G9}. For a completely different
subsequent proof, see Sadasue \cite{Sa}.

The original proof of the Hermite expansion for $K$ was in \cite{G1} and
involved the stochastic analysis described here. A purely finite-dimensional
proof of the Hermite theorem for $K$ was then given by Hijab \cite{Hi1,Hi2}
and Driver \cite{Dr}. Meanwhile, Gross's results for the Hermite expansion
motivated my own development of the Segal-Bargmann transform for $K$ in
\cite{H1}. The proofs in \cite{H1} are purely finite-dimensional. Later, Gross
and Malliavin in \cite{GM} showed how to understand the Segal-Bargmann
transform for $K$ in the way described here, connecting it with the
infinite-dimensional linear theory.

P. Biane \cite{Bi} has given a generalization of the Gross-Malliavin result to
the setting of free probability theory. M. Gordina \cite{Go1,Go2} has studied
versions of the Driver-Gross theorem (Theorem \ref{dg.thm}) in the setting of
infinite-dimensional Lie groups. T. Deck has given a group analog of the
notion of Hida distributions in white noise analysis \cite{De}.

\subsection{The Yang-Mills interpretation}

In \cite{Wr}, K. Wren considers the problem of canonical quantization of
Yang-Mills theory on a spacetime cylinder, using a method proposed by N.
Landsman \cite{L1}. (See also Chapter IV.3.8 of \cite{L2}.) Wren's
calculations strongly suggest a close relationship between this Yang-Mills
example and the $K$-invariant form of the Segal-Bargmann transform for a
compact Lie group. The paper \cite{DH} investigates this relationship further.
(See also the expository paper \cite{H7}.) Driver and I make a modification of
the Gross-Malliavin results that fits with the desired Yang-Mills
interpretation. (See also \cite{A,AHS,Lo} for other uses of the Segal-Bargmann
transform for $K$ in connection with quantized gauge theories.)

The Yang-Mills problem considered is the simplest non-trivial case, namely
Yang-Mills theory on a spacetime cylinder. We take $K$ as our structure group.
So time is a line, and space is a circle, which we think of as the interval
$\left[  0,1\right]  $ with ends identified. We consider first
\textit{classical} Yang-Mills theory, using the temporal gauge. In the
temporal gauge, the Yang-Mills equations may be considered as an
infinite-dimensional Hamiltonian system whose \textbf{configuration space} is
the space of square-integrable Lie algebra-valued 1-forms on $\left[
0,1\right]  .$ Since space is one-dimensional, we identify the configuration
space with the space of square-integrable Lie algebra-valued functions. That
is, our configuration space is $H_{\mathbb{R}}=L^{2}(\left[  0,1\right]
;\frak{k}).$ The associated \textbf{phase space} for the classical Yang-Mills
theory is then the cotangent bundle of $H_{\mathbb{R}},$ which may be
identified with $H_{\mathbb{C}}=L^{2}(\left[  0,1\right]  ;\frak{k}%
_{\mathbb{C}}).$

The \textbf{based gauge group} for this theory is the set of maps $x$ of the
space manifold $\left[  0,1\right]  $ into the structure group $K,$ such that
$x_{0}=x_{1}=e.$ If (for technical reasons) we limit ourselves to maps with
one derivative in $L^{2}$, then the based gauge group is nothing but the loop
group $\mathcal{L}\left(  K\right)  .$ The based gauge group acts on
$H_{\mathbb{R}}$ in the usual way in gauge theory, namely by
\[
\left(  l\cdot A\right)  _{\tau}=l_{\tau}A_{\tau}l_{\tau}^{-1}-\frac{dl}%
{d\tau}l_{\tau}^{-1},\quad l\in\mathcal{L}\left(  K\right)  ,\,A\in\bar
{H}_{\mathbb{R}}.
\]
This is precisely the action (\ref{loop.act}) from the previous subsection,
but now with a different interpretation. In the classical Yang-Mills theory,
one is supposed to \textbf{reduce} by the action of the gauge group. This
means that we replace the phase space $H_{\mathbb{C}}=T^{\ast}(H_{\mathbb{R}%
})$ by $T^{\ast}(H_{\mathbb{R}}/\mathcal{L}\left(  K\right)  ).$ So $T^{\ast
}(H_{\mathbb{R}})$ is the unreduced phase space and $T^{\ast}(H_{\mathbb{R}%
}/\mathcal{L}\left(  K\right)  )$ is the reduced or physical phase space. For
a more detailed explanation of this reduction, see \cite{DH,H7,L2}.

It is not difficult to show that two elements $A_{1}$ and $A_{2}$ of
$H_{\mathbb{R}}$ are gauge-equivalent (i.e. there exists $l\in\mathcal{L}%
\left(  K\right)  $ with $l\cdot A_{1}=A_{2}$) if and only if $h\left(
A_{1}\right)  =h\left(  A_{2}\right)  .$ Here $h\left(  \cdot\right)  $ is the
holonomy map of (\ref{holonomy.form}), with the terminology motivated by this
the Yang-Mills interpretation. So the gauge equivalence classes are labeled by
the holonomy $h\left(  A\right)  \in K.$ This means that
\begin{equation}
H_{\mathbb{R}}/\mathcal{L}\left(  K\right)  \cong K.\label{reduce.k}%
\end{equation}

I will not discuss the dynamics of the classical Yang-Mills theory. (For the
dynamics see \cite{DH}.) Instead I will consider quantizing the theory. The
idea is to consider first quantizing the unreduced theory, obtaining a
position Hilbert space, a Segal-Bargmann space, and a Segal-Bargmann transform
corresponding to the configuration space $H_{\mathbb{R}}.$ Having done that we
then want to reduce appropriately by the gauge group $\mathcal{L}\left(
K\right)  ,$ which in light of (\ref{reduce.k}) ought to give us a position
Hilbert space, Segal-Bargmann space, and Segal-Bargmann transform for $K.$ The
set-up of the last section \textit{almost} gives us what we want: we think of
$L^{2}(\bar{H}_{\mathbb{R}},P_{t})$ as our position Hilbert space, and then we
``reduce'' by restricting to the functions in $L^{2}(\bar{H}_{\mathbb{R}%
},P_{t})$ that are invariant under the action of the loop group. By Theorem
\ref{loop.thm} (transferred to $\bar{H}_{\mathbb{R}}$), the loop-invariant
functions are just functions of the holonomy, and by Theorem \ref{gm.thm}
these map under the Segal-Bargmann transform to functions of the complex holonomy.

Unfortunately, this set-up is not quite right from the point of view of
quantization, because the loop group action on $L^{2}(\bar{H}_{\mathbb{R}%
},P_{t})$ is not unitary. That is, since the action of $\mathcal{L}\left(
K\right)  $ on $\bar{H}_{\mathbb{R}}$ leaves the measure $P_{t}$ only
quasi-invariant, but not invariant, the $L^{2}$-norm of $f\left(  l\cdot
A\right)  $ need not equal the $L^{2}$-norm of $f\left(  A\right)  .$ Since in
quantization theory the symmetries of the problem are supposed to act by
unitary transformations, we have a problem.

To fix this problem, Driver and I considered as our position Hilbert space
$L^{2}(\bar{H}_{\mathbb{R}},P_{s}),$ where the variance parameter $s$ is
large. Formally, as $s\rightarrow\infty,$ $P_{s}$ converges to the
(non-existent) Lebesgue measure on $\bar{H}_{\mathbb{R}},$ so that the loop
group action becomes more and more nearly unitary. We then want to restrict
attention to the loop-invariant subspace and apply the Segal-Bargmann
transform. However, if we simply use the ordinary transform $B_{s},$ it will
not make sense in the $s\rightarrow\infty$ limit. So we introduce a
``two-parameter transform'' $B_{s,t}.$ This transform is given by the same
formula as the time $t$ transform $B_{t},$ but defined on the time $s$ Hilbert
space $L^{2}(\bar{H}_{\mathbb{R}},P_{s}).$ The target Hilbert space is
$\mathcal{H}L^{2}(\bar{H}_{\mathbb{R}},M_{s,t}),$ where $M_{s,t}$ is an
appropriately defined Gaussian measure on $\bar{H}_{\mathbb{R}}.$ (See
\cite{DH,H5} for details.)

\begin{theorem}
\label{st.thm}The following diagram commutes and all maps are unitary.
\[%
\begin{array}
[c]{ccc}%
L^{2}(\bar{H}_{\mathbb{R}},P_{s})^{\mathcal{L}\left(  K\right)  } &
\underleftrightarrow{B_{s,t}} & \mathcal{H}L^{2}(\bar{H}_{\mathbb{C}}%
,M_{s,t}^{{}})^{\mathcal{L}\left(  K\right)  }\\
\updownarrow\,\circ h &  & \updownarrow\,\circ h_{\mathbb{C}}\\
L^{2}(K,\rho_{s})\quad & \underleftrightarrow{B_{s,t}} & \mathcal{H}%
L^{2}(K_{\mathbb{C}},\mu_{s,t})\quad
\end{array}
\]
Here the superscript $\mathcal{L}\left(  K\right)  $ denotes functions
invariant under the action of the loop group. On the top row, $B_{s,t}$ is the
modified Segal-Bargmann transform for the linear space $\bar{H}_{\mathbb{R}}.$
On the bottom row, $B_{s,t}$ is the modified Segal-Bargmann transform for $K,$
given by
\[
B_{s,t}f=\text{ analytic continuation of }e^{t\Delta_{K}/2}f.
\]

It makes sense to take the limit $s\rightarrow\infty$ in the bottom row, and
the result is the $K$-invariant Segal-Bargmann transform
\[
C_{t}:L^{2}(K,dx)\rightarrow\mathcal{H}L^{2}(K_{\mathbb{C}},\nu_{t}),
\]
as described in Section 3.
\end{theorem}

To summarize in words: if one takes the modified Segal-Bargmann transform
$B_{s,t}$ for $\bar{H}_{\mathbb{R}},$ restricts to the loop-invariant
subspace, and lets $s$ tend to infinity, the result is the $K$-invariant
Segal-Bargmann transform for $K.$ This is consistent with the results of Wren
\cite{Wr}, who approached the problem in a different way. The proofs in
\cite{DH} rely on the Hermite decomposition (chaos expansion) for $L^{2}%
(\bar{H}_{\mathbb{R}},P_{s}).$

Let me close this section by considering these results from the point of view
of geometric quantization. (See the discussion in \cite[Sect. 8]{H7}.) The
invariant form of Segal-Bargmann transform for $\mathbb{R}^{d}$ can be
obtained by geometric quantization (the $\mathbb{R}^{d}$ case of what is
described in Section 3.2). If one lets $d\rightarrow\infty,$ one gets roughly
$\mathcal{H}L^{2}(\bar{H}_{\mathbb{C}},M_{s,t}),$ where the extra parameter
$s\gg t$ is a necessary regularization. So $\mathcal{H}L^{2}(\bar
{H}_{\mathbb{C}},M_{s,t})$ may be thought of as the result of applying
geometric quantization to the phase space $H_{\mathbb{C}}.$ If one then
reduces by $\mathcal{L}\left(  K\right)  ,$ the result (by Theorem
\ref{st.thm}) is naturally identifiable with $\mathcal{H}L^{2}\left(
K_{\mathbb{C}},\nu_{t}\right)  .$ On the other hand, if we \textit{first}
reduce by $\mathcal{L}\left(  K\right)  $ to get the phase space $T^{\ast
}(H_{\mathbb{R}}/\mathcal{L}\left(  K\right)  )=T^{\ast}(K),$ and
\textit{then} perform geometric quantization, Section 3.2 tells us that the
result is again $\mathcal{H}L^{2}\left(  K_{\mathbb{C}},\nu_{t}\right)  .$ So
we may say that in this instance ``quantization commutes with reduction.''
That is, there is a natural \textit{unitary} correspondence between the
Hilbert space obtained by first doing geometric quantization and then reducing
and the Hilbert space obtained the other way around.

The question of how quantization relates to reduction is an old and important
one. V. Guillemin and S. Sternberg \cite{GStern} sparked a recent surge of
interest in the problem by considering it in the context of geometric
quantization of compact K\"{a}hler manifolds. They consider quantization
without the half-forms, and they show that under certain regularity
assumptions there is a natural one-to-one linear correspondence between the
Hilbert space obtained by first quantizing and then reducing and the one
obtained by first reducing and then quantizing. However, they do not prove
that this correspondence is unitary, and indeed it seems unlikely to be
unitary in general. By contrast, in the Yang-Mills example considered here, we
have included the half-forms in the quantization, and we have obtained a
\textit{unitary} correspondence.

\begin{problem}
Give general conditions under which quantization of K\"{a}hler manifolds, with
or without the half-form correction, commutes \textbf{unitarily} with reduction.
\end{problem}

\noindent I believe that one is more likely to get a unitary correspondence if
one includes the half-forms, but even then I do not expect unitarity in general.

I should add that in the setting of compact K\"{a}hler manifolds, as
considered in \cite{GStern}, the Hilbert spaces obtained are
finite-dimensional. In that setting an important consequence of the
Guillemin-Sternberg correspondence (unitary or not) is that the dimension of
the space obtained in the ``first quantize and then reduce'' approach is the
same as the dimension of the space obtained in the ``first reduce then
quantize'' approach. For non-compact K\"{a}hler manifolds the Hilbert spaces
are typically infinite-dimensional.

\section{Proof sketches}

\subsection{The Segal-Bargmann transform for $K$}

I present here an argument for Theorem \ref{sb.thm2}, as given (in slightly
greater generality) in \cite{H5}. The method of proving isometricity is
essentially that proposed by T. Thiemann in \cite[Sect. 2.3]{T}.

Since $\rho_{t}\left(  x\right)  $ is the heat kernel at the identity, we have
that for any function $\phi$ on $K$%
\begin{equation}
\int_{K}\phi\left(  x\right)  \rho_{t}\left(  x\right)  \,dx=\left(
e^{t\Delta_{K}/2}\phi\right)  \left(  e\right)  .\label{heat.int1}%
\end{equation}
Similarly, if $\psi$ is a function on $K_{\mathbb{C}}$ then
\begin{equation}
\int_{K_{\mathbb{C}}}\psi\left(  g\right)  \mu_{t}\left(  g\right)
\,dg=\left(  e^{t\Delta_{K_{\mathbb{C}}}/4}\psi\right)  \left(  e\right)
.\label{heat.int2}%
\end{equation}
Thus the isometricity of the Segal-Bargmann transform is equivalent to the
statement that for all $f\in L^{2}(K,\rho_{t})$ we have
\begin{equation}
e^{t\Delta_{K}/2}\left(  \bar{f}\,f\right)  \left(  e\right)  =e^{t\Delta
_{K_{\mathbb{C}}}/4}\left(  \overline{e^{t\Delta_{K}/2}f}\,e^{t\Delta_{K}%
/2}f\right)  \left(  e\right)  .\label{norm1}%
\end{equation}
On the right we have implicitly analytically continued $e^{t\Delta_{K}/2}f$ to
$K_{\mathbb{C}}.$

Let us assume that $f$ itself admits an analytic continuation to
$K_{\mathbb{C}}.$ (The space of such $f$'s is dense.) Now, $\Delta_{K}=\sum
X_{k}^{2}$, regarded as a left-invariant differential operator on
$K_{\mathbb{C}},$ commutes with complex conjugation and with analytic
continuation. Thus
\[
\overline{e^{t\Delta_{K}/2}f}=e^{t\Delta_{K}/2}\bar{f}.
\]
Note that on the left, we are first applying $e^{t\Delta_{K}/2},$ then
analytically continuing, and then taking the complex conjugate. On the right
we are first analytically continuing, then taking the complex conjugate and
then applying $e^{t\Delta_{K}/2}.$

Next consider the operators
\begin{align*}
Z_{k}  & =\frac{1}{2}\left(  X_{k}-iJX_{k}\right) \\
\bar{Z}_{k}  & =\frac{1}{2}\left(  X_{k}+iJX_{k}\right)  ,
\end{align*}
which reduce in the case $K_{\mathbb{C}}=\mathbb{C}^{d}$ to $\partial/\partial
z_{k}$ and $\partial/\partial\bar{z}_{k}.$ On the holomorphic function $f$ we
have $Z_{k}f=X_{k}f$ and $\bar{Z}_{k}f=0,$ and on the anti-holomorphic
function $\bar{f},$ $Z_{k}\bar{f}=0$ and $\bar{Z}_{k}f=X_{k}f$. It follows
that
\[
\overline{e^{t\Delta_{K}/2}f}\,e^{t\Delta_{K}/2}f=e^{t\sum Z_{k}^{2}%
/2}e^{t\sum\bar{Z}_{k}^{2}/2}\left(  \bar{f}\,f\right)  .
\]
So the desired norm equality becomes
\begin{equation}
e^{t\Delta_{K}/2}\left(  \bar{f}\,f\right)  \left(  e\right)  =e^{t\Delta
_{K_{\mathbb{C}}}/4}e^{t\sum Z_{k}^{2}/2}e^{t\sum\bar{Z}_{k}^{2}/2}\left(
\bar{f}\,f\right)  \left(  e\right)  .\label{norm2}%
\end{equation}

Now, a holomorphic vector field $Z_{k}$ automatically commutes with an
anti-holomorphic vector field $\bar{Z}_{l}$ (or calculate this directly). Thus
the second and third exponents on the right of (\ref{norm2}) may be combined.
The exponent that results is
\[
\frac{t}{2}\sum_{k=1}^{\dim\frak{k}}\left(  Z_{k}^{2}+\bar{Z}_{k}^{2}\right)
=\frac{t}{4}\sum_{k=1}^{\dim\frak{k}}\left(  X_{k}^{2}-(JX_{k})^{2}\right)  .
\]
This is a constant times the Casimir operator for $K_{\mathbb{C}},$ which is
bi-invariant and therefore commutes with the left-invariant operator
$\Delta_{K_{\mathbb{C}}}.$ So in the end all three exponents on the right in
(\ref{norm2}) may be combined. It thus suffices to have the sum of the three
exponents on the right in (\ref{norm2}) equal to the exponent on the left. So
we need
\begin{equation}
\frac{t}{2}\sum_{k=1}^{\dim\frak{k}}X_{k}^{2}=\frac{t}{4}\sum_{k=1}%
^{\dim\frak{k}}X_{k}^{2}+\frac{t}{4}\sum_{k=1}^{\dim\frak{k}}(JX_{k}%
)^{2}+\frac{t}{4}\sum_{k=1}^{\dim\frak{k}}\left(  X_{k}^{2}-(JX_{k}%
)^{2}\right)  ,\label{norm3}%
\end{equation}
which is true!

It is not difficult to make this argument rigorous on a dense subspace of
$L^{2}(K,\rho_{t}),$ thus establishing the isometricity of the Segal-Bargmann
transform $B_{t}.$ Using Proposition \ref{structure.prop} (see also the
appendix of \cite{H5}), the surjectivity can be reduced to two cases:
$K=\mathbb{R}^{d}$ and $K$ compact. In the $\mathbb{R}^{d}$ case, one shows
that the image of $B_{t}$ contains all holomorphic polynomials, which are
known \cite[Sect. 1b]{B} to be dense in $\mathcal{H}L^{2}(\mathbb{C}^{d}%
,\mu_{t}).$ In the case $K$ compact, it is easily seen that the image of
$B_{t}$ contains all the matrix entries for finite-dimensional holomorphic
representations of $K_{\mathbb{C}}.$ Using the Peter-Weyl theorem and the
``averaging lemma'' \cite[Lem. 11]{H1} one can show that these holomorphic
matrix entries are dense in $\mathcal{H}L^{2}(K_{\mathbb{C}},\mu_{t})$ and
thus that $B_{t}$ is surjective.

\subsection{The Taylor map}

The argument used here is essentially that of \cite{Dr} and \cite{DG}, except
that I give it in ``exponentiated form.'' If $F\in\mathcal{H}L^{2}%
(K_{\mathbb{C}},\mu_{t})$ then by (\ref{heat.int2})
\begin{align*}
\left\|  F\right\|  _{L^{2}\left(  K_{\mathbb{C}},\mu_{t}\right)  }^{2}  &
=\int_{K_{\mathbb{C}}}\left|  F\left(  g\right)  \right|  ^{2}\mu_{t}\left(
g\right)  \,dg\\
& =\left.  e^{t\Delta_{K_{\mathbb{C}}}/4}\left(  \bar{F}\left(  g\right)
F\left(  g\right)  \right)  \right|  _{g=e}.
\end{align*}
But in the notation of the previous subsection we have
\[
\frac{\Delta_{K_{\mathbb{C}}}}{4}=\sum_{k=1}^{\dim\frak{k}}\bar{Z}_{k}Z_{k}.
\]
Since $Z_{k}$ annihilates the anti-holomorphic function $\bar{F}\left(
g\right)  $ and $\bar{Z}_{k}$ annihilates the holomorphic function $F\left(
g\right)  ,$ when applying $\Delta_{K_{\mathbb{C}}}$ to $\bar{F}\left(
g\right)  F\left(  g\right)  ,$ only the ``cross terms'' survive. That is,
\[
\frac{\Delta_{K_{\mathbb{C}}}}{4}\left(  \bar{F}\left(  g\right)  F\left(
g\right)  \right)  =\sum_{k=1}^{\dim\frak{k}}\bar{Z}_{k}\bar{F}\left(
g\right)  Z_{k}F\left(  g\right)  .
\]
Also, $\bar{Z}_{k}\bar{F}=X_{k}\bar{F}$ and $Z_{k}F=X_{k}F.$ Thus
\[
\frac{\Delta_{K_{\mathbb{C}}}}{4}\left(  \bar{F}\left(  g\right)  F\left(
g\right)  \right)  =\sum_{k=1}^{\dim\frak{k}}X_{k}\bar{F}\left(  g\right)
X_{k}F\left(  g\right)  .
\]

More generally,
\[
\left(  \frac{\Delta_{K_{\mathbb{C}}}}{4}\right)  ^{n}\left(  \bar{F}\left(
g\right)  F\left(  g\right)  \right)  =\sum_{k_{1},\cdots,k_{n}=1}%
^{\dim\frak{k}}X_{k_{1}}\cdots X_{k_{n}}\bar{F}\left(  g\right)  X_{k_{1}%
}\cdots X_{k_{n}}F\left(  g\right)  .
\]
Thus at least formally we have
\[
\left.  e^{t\Delta_{K_{\mathbb{C}}}/4}\left(  \bar{F}\left(  g\right)
F\left(  g\right)  \right)  \right|  _{g=e}=\sum_{n=0}^{\infty}\frac{t^{n}%
}{n!}\sum_{k_{1},\cdots,k_{n}=1}^{\dim\frak{k}}X_{k_{1}}\cdots X_{k_{n}}%
\bar{F}\left(  e\right)  X_{k_{1}}\cdots X_{k_{n}}F\left(  e\right)  .
\]
This is just the basis-dependent statement of the isometricity of the Taylor map.

Note that we do not need to have any commutativity for this argument to work,
in contrast to the argument for the isometricity of the Segal-Bargmann
transform. This observation is the basis of the results of Driver-Gross
\cite{DG} (Theorem \ref{dg.thm} of Section 2.2).

Of course the above argument is purely formal. In the compact group case it is
not hard to justify it on a dense subspace of $\mathcal{H}L^{2}(K_{\mathbb{C}%
},\mu_{t}).$ Then one has to prove that if $K_{\mathbb{C}}$ is simply
connected, the Taylor map is \textit{onto} $J_{t}^{0}.$ Suppose then that
$\xi$ is an element of $J_{t}^{0}.$ Then using a construction of Gross
\cite[Sect. 8]{G2}, it is not too difficult to produce a holomorphic function
$F$ whose derivatives at the identity are given by $\xi.$ The construction of
$F$ is a ``Taylor expansion along paths,'' with the simple-connectedness use
to show independence of path. (See \cite[Sects. 5 and 6]{Dr}.) So to complete
the proof, one merely needs to show that the resulting function $F$ is in
$\mathcal{H}L^{2}(K_{\mathbb{C}},\mu_{t}).$ Certainly this \textit{ought} to
be the case, since the norm of $F$ in $\mathcal{H}L^{2}(K_{\mathbb{C}},\mu
_{t})$ should be equal to the norm of $\xi$ in $J_{t}^{0}.$ However, the proof
of isometricity of the Taylor map only works if one knows ahead of time that
$F$ in $\mathcal{H}L^{2}(K_{\mathbb{C}},\mu_{t}).$ Instead, then, one
estimates the growth of $F$ and gets bounds (Theorem \ref{bounds.thm})
sufficient to show that $F\in\mathcal{H}L^{2}(K_{\mathbb{C}},\mu_{s})$ for all
$s<t$. Then using the Taylor isometry at time $s, $ one shows that the norm of
$F$ at time $s$ is bounded as $s$ increases to $t.$ It is then possible to
show that the norm of $F$ at time $t$ is finite as well. See \cite[Thm.
5.7]{Dr}.

\subsection{The Hermite expansion}

Although the Hermite expansion is just the composition of the Segal-Bargmann
transform and the Taylor map, it is illuminating to give a direct proof of its
isometricity that does not explicitly involve the complex group $K_{\mathbb{C}%
}.$ As in (\ref{heat.int1}) in Section 5.1, the norm of a function $f\in
L^{2}(K,\rho_{t})$ may be computed as
\[
\left\|  f\right\|  _{L^{2}\left(  K,\rho_{t}\right)  }^{2}=e^{t\Delta_{K}%
/2}\left(  \bar{f}\,f\right)  \left(  e\right)  .
\]
If we formally expand the heat operator $e^{t\Delta_{K}/2}$ in powers of the
Laplacian, then we will have to apply powers of the Laplacian to the product
$\bar{f}\,f.$ This will involve applying the product rule repeatedly. The use
of the product rule can be organized as follows:
\[
X_{k}\left(  \bar{f}\,f\right)  \left(  e\right)  =\left(  X_{k}+Y_{k}\right)
\left.  \left(  \bar{f}\left(  x\right)  f\left(  y\right)  \right)  \right|
_{x=y=e},
\]
where $X_{k}$ means a derivative in the $x$-variable and $Y_{k}$ means the
same derivative but in the $y$-variable. Applying this idea repeatedly we see
that
\[
\Delta_{K}\left(  \bar{f}\,f\right)  \left(  e\right)  =\sum_{k=1}%
^{\dim\frak{k}}\left(  X_{k}+Y_{K}\right)  ^{2}\left.  \left(  \bar{f}\left(
x\right)  f\left(  y\right)  \right)  \right|  _{x=y=e}
\]
and more generally
\[
\Delta_{K}^{n}\left(  \bar{f}\,f\right)  \left(  e\right)  =\left(  \sum
_{k=1}^{\dim\frak{k}}\left(  X_{k}+Y_{k}\right)  ^{2}\right)  ^{n}\left.
\left(  \bar{f}\left(  x\right)  f\left(  y\right)  \right)  \right|
_{x=y=e}.
\]

Thus formally we may write
\begin{equation}
e^{t\Delta_{K}/2}\left(  \bar{f}\,f\right)  \left(  e\right)  =\exp\left(
\frac{t}{2}\sum_{k=1}^{\dim\frak{k}}\left(  X_{k}+Y_{k}\right)  ^{2}\right)
\left.  \left(  \bar{f}\left(  x\right)  f\left(  y\right)  \right)  \right|
_{x=y=e}.\label{exp.eq}%
\end{equation}
Since all the $X_{k}$'s automatically commute with the $Y_{l}$'s (since they
act on different variables) we have
\begin{equation}
\sum_{k=1}^{\dim\frak{k}}\left(  X_{k}+Y_{k}\right)  ^{2}=\sum_{k=1}%
^{\dim\frak{k}}X_{k}^{2}+\sum_{k=1}^{\dim\frak{k}}Y_{k}^{2}+2\sum_{k=1}%
^{\dim\frak{k}}X_{k}Y_{k}.\label{xkyk}%
\end{equation}
But $\Sigma X_{k}^{2}$ is a bi-invariant operator on $K,$ and so it commutes
with each $X_{l}.$ Thus in fact all three terms on the right side of
(\ref{xkyk}) commute with each other. So we may (formally) factor the
exponential right in (\ref{exp.eq}) into a product of three exponentials. If
we let the terms with $X_{k}^{2}$ and $Y_{k}^{2}$ act first, then one will act
only on $\bar{f}\left(  x\right)  $ and one will act only on $f\left(
y\right)  .$ So we get
\[
e^{t\Delta_{K}/2}\left(  \bar{f}\,f\right)  \left(  e\right)  =\exp\left(
t\sum_{k=1}^{\dim\frak{k}}X_{k}Y_{k}\right)  \left.  \left(  \left[
e^{t\Delta_{K}/2}\bar{f}\left(  x\right)  \right]  \left[  e^{t\Delta_{K}%
/2}f\left(  y\right)  \right]  \right)  \right|  _{x=y=e}.
\]

Now we may expand out the exponential involving $X_{k}Y_{k}$ to give
\begin{align*}
\left\|  f\right\|  _{L^{2}\left(  K,\rho_{t}\right)  }^{2}  & =e^{t\Delta
_{K}/2}\left(  \bar{f}\,f\right)  \left(  e\right) \\
& =\sum_{n=0}^{\infty}\frac{t^{n}}{n!}\sum_{k_{1},\cdots,k_{n}}\left.
X_{k_{1}}\cdots X_{k_{n}}e^{t\Delta_{K}/2}\bar{f}\left(  x\right)  Y_{k_{1}%
}\cdots Y_{k_{n}}e^{t\Delta_{K}/2}f\left(  y\right)  \right|  _{x=y=e}.
\end{align*}
Once we evaluate everything at $e$ there is no need to use different letters
for the $x$-derivatives and the $y$-derivatives and we get simply
\[
\left\|  f\right\|  ^{2}=\sum_{n=0}^{\infty}\frac{t^{n}}{n!}\sum_{k_{1}%
,\cdots,k_{n}=1}^{\dim\frak{k}}\left|  X_{k_{1}}\cdots X_{k_{n}}e^{t\Delta
_{K}/2}f\left(  e\right)  \right|  ^{2}.
\]
This is nothing but the statement of the isometricity of the Hermite expansion
from Theorem \ref{hermite.thm2}.

If $K$ is simply connected, then one can argue directly for the surjectivity
of the Hermite expansion as in \cite{Hi1,Hi2}. So we let $\xi$ be an element
of $J_{t}^{0}.$ We want to construct a function $f$ such that $\xi$ encodes
all of the derivatives of $e^{t\Delta_{K}/2}f$ at the identity. Using ``Taylor
expansion along paths'' as in the last subsection we can produce a function
$F$ whose derivatives at the identity are given by $\xi.$ (This is where the
simple connectedness of $K$ is used.) We think of $F$ now as simply a function
on $K$, even though $F$ extends to a holomorphic function on $K_{\mathbb{C}}.$
We have to prove that there exists a function $f\in L^{2}(K,\rho_{t})$ such
that $F=e^{t\Delta_{K}/2}f.$ Hijab argues first that for all $s<t$ there
exists $f_{s}$ with $e^{s\Delta_{K}/2}f_{s}=F.$ Then using the isometricity of
the Hermite expansion at time $s,$ he argues that the norm of $f_{s}$ remains
bounded as $s$ increases to $t,$ in which case one may show that
$f=\lim_{s\uparrow t}f_{s}$ exists with $e^{t\Delta_{K}/2}f=F.$


\begin{thebibliography}{99}
\bibitem[AHS]{AHS}S. Albeverio, B. Hall, and A. Sengupta, The Segal-Bargmann
transform for two-dimensional Euclidean quantum Yang-Mills, \textit{Infinite
Dimensional Anal. Quantum Prob.} \textbf{2} (1999), 27-49.

\bibitem[A] {A}A. Ashtekar, J. Lewandowski, D. Marolf, J. Mour\~{a}o, and T.
Thiemann, Coherent state transforms for spaces of connections, \textit{J.
Funct. Anal.} \textbf{135} (1996) 519-551.

\bibitem[BSZ] {BSZ}J. Baez, I. Segal, and Z. Zhou, ``Introduction to
Algebraic and Constructive Quantum Field Theory,'' Princeton Univ. Press,
Princeton, NJ, 1992.

\bibitem[B] {B}V. Bargmann, On a Hilbert space of analytic functions and an
associated integral transform, Part I, \textit{Comm. Pure Appl. Math.}
\textbf{14} (1961), 187-214.

\bibitem[Bi] {Bi}P. Biane, Segal-Bargmann transform, functional calculus on
matrix spaces and the theory of semi-circular and circular systems, \textit{J.
Funct. Anal.} \textbf{144} (1997), 232--286.

\bibitem[De]{De}  T. Deck, Hida distributions on compact Lie groups, \textit{%
Infinite Dimensional Anal. Quantum Prob.} \textbf{3} (2000), 1-26.

\bibitem[Dr] {Dr}B. Driver, On the Kakutani-It\^{o}-Segal-Gross and
Segal-Bargmann-Hall isomorphisms, \textit{J. Funct. Anal.} \textbf{133}
(1995), 69-128.

\bibitem[DG] {DG}B. Driver and L. Gross, Hilbert spaces of holomorphic
functions on complex Lie groups. \textit{In} ``New trends in stochastic
analysis. Proceedings of a Taniguchi international workshop,'' (K. Elworthy,
S. Kusuoka, and I. Shigekawa, Eds.) , pp. 76-106. World Scientific, Singapore, 1997.

\bibitem[DH] {DH}B. Driver and B. Hall, Yang-Mills theory and the
Segal-Bargmann transform, \textit{Commun. Math. Phys.} \textbf{201} (1999), 249-290.

\bibitem[E] {E}M. Engli\v{s}, Asymptotic behaviour of reproducing kernels of
weighted Bergman spaces, \textit{Trans. Amer. Math. Soc.} \textbf{349} (1997), 3717--3735.

\bibitem[F-J] {F-J}M. Flensted-Jensen, Spherical functions on a real
semisimple Lie group. A method of reduction to the complex case, \textit{J.
Funct. Anal.} \textbf{30} (1978), 106-146.

\bibitem[F] {F}G. Folland, ``Harmonic analysis in phase space,'' Princeton
Univ. Press, Princeton, N.J., 1989.

\bibitem[FS] {FS}G. Folland and A. Sitaram, The uncertainty principle: a
mathematical survey, \textit{J. Fourier Anal. Appl.} \textbf{3} (1997), 207--238.

\bibitem[Go1] {Go1}M. Gordina, Holomorphic functions and the heat kernel
measure on an infinite dimensional complex orthogonal group, \textit{Potential
Anal.}, to appear

\bibitem[Go2] {Go2}M. Gordina, Heat kernel analysis and Cameron-Martin
subgroup for infinite dimensional groups, \textit{J. Funct. Anal.}
\textbf{171} (2000), 192-232.

\bibitem[G1] {G1}L. Gross, Abstract Wiener spaces, \textit{in} ``Proceedings
of the Fifth Berkeley Symposium on Mathematical Statistics and Probablility,''
Vol. II, Univ. of California Press, 1967.

\bibitem[G2] {G2}L. Gross, Uniqueness of ground states for Schr\"{o}dinger
operators over loop groups, \textit{J. Funct. Anal.} \textbf{112} (1993) 373-441.

\bibitem[G3] {G3}L. Gross, Analysis on loop groups, \textit{in} ``Stochastic
analysis and applications in physics (Funchal, 1993)'', pp. 99--118, NATO Adv.
Sci. Inst. Ser. C Math. Phys. Sci., 449, Kluwer Acad. Publ., Dordrecht, 1994.

\bibitem[G4] {G4}L. Gross, Harmonic functions on loop groups. \textit{S\'{e}%
minaire Bourbaki} Vol. 1997/98. Ast\'{e}risque No. 252, (1998), Exp. No. 846,
5, 271--286.

\bibitem[G5] {G5}L. Gross, The homogeneous chaos over compact Lie groups,
\textit{in} ``Stochastic processes: A Festschrift in Honour of Gopinath
Kallianpur'' (S. Cambanis \textit{et al}., Eds.), pp. 117--123, Springer, New
York, 1993.

\bibitem[G6] {G6}L. Gross, Harmonic analysis for the heat kernel measure on
compact homogeneous spaces, \textit{in} ``Stochastic analysis on
infinite-dimensional spaces (Baton Rouge, LA, 1994)'' (H. Kunita and H.-H.
Kuo, Eds.), pp. 99--110, Pitman Res. Notes Math. Ser., 310, Longman Sci.
Tech., Harlow, 1994.

\bibitem[G7] {G7}L. Gross, Some norms on universal enveloping algebras,
\textit{Canad. J. Math.} \textbf{50} (1998), 356--377.

\bibitem[G8] {G8}L. Gross, A local Peter-Weyl theorem, \textit{Trans. Amer.
Math. Soc.} \textbf{352} (2000), 413--427.

\bibitem[G9] {G9}L. Gross, Heat kernel analysis on Lie groups, preprint.

\bibitem[GM] {GM}L. Gross and P. Malliavin, Hall's transform and the
Segal-Bargmann map, \textit{in ``}It\^{o}'s stochastic calculus and
probability theory,''\ (M.\textit{\ }Fukushima, N. Ikeda, H. Kunita, and S.
Watanabe, Eds.), pp. 73-116. Springer-Verlag, Berlin/New York, 1996.

\bibitem[GStenz] {GStenz}V. Guillemin and M. Stenzel, Grauert tubes and the
homogeneous Monge-Amp\`{e}re equation, \textit{J. Differential Geom.}
\textbf{34} (1991), 561--570.

\bibitem[GStern] {GStern}V. Guillemin and S. Sternberg, Geometric
quantization and multiplicities of group representations, \textit{Invent.
Math.} \textbf{67} (1982), 515--538.

\bibitem[H1] {H1}B. Hall, The Segal-Bargmann ``coherent state'' transform for
compact Lie groups, \textit{J. Funct. Anal.} \textbf{122} (1994), 103-151.

\bibitem[H2] {H2}B. Hall, The inverse Segal-Bargmann transform for compact
Lie groups, \textit{J. Funct. Anal.}, \textbf{143} (1997), 98-116.

\bibitem[H3] {H3}B. Hall, Phase space bounds for quantum mechanics on a
compact Lie group, \textit{Comm. Math. Phys.}, \textbf{184} (1997), 233-250.

\bibitem[H4] {H4}B. Hall, Quantum mechanics in phase space, \textit{in}
``Perspectives on quantization'' (L. Coburn and M. Rieffel, Eds), pp. 47--62,
Contemporary Mathematics, Vol. 214, Amer. Math. Soc., Providence, RI, 1998.

\bibitem[H5] {H5}B. Hall, A new form of the Segal-Bargmann transform for Lie
groups of compact type, \textit{Canad. J. Math.} \textbf{51} (1999), 816-834.

\bibitem[H6] {H6}B. Hall, Holomorphic methods in analysis and mathematical
physics, \textit{to appear in} ``First Summer School in Analysis and
Mathematical Physics, Cuernavaca, Mexico,'' (S. P\'{e}rez Esteva and C.
Villegas Blas, Eds.), Contemporary Mathematics, Amer. Math. Soc., Providence,
RI, 2000. [htttp://xxx.lanl.gov, quant-ph/9912054]

\bibitem[H7] {H7}B. Hall, Coherent states, Yang-Mills theory, and reduction,
preprint. [htttp://xxx.lanl.gov, quant-ph/9911052]

\bibitem[H8] {H8}B. Hall, Geometric quantization and the generalized
Segal-Bargmann transform, in preparation.

\bibitem[HS] {HS}B. Hall and A. Sengupta, The Segal-Bargmann transform for
path-groups, \textit{J. Funct. Anal.} \textbf{152} (1998), 220-254.

\bibitem[He] {He}S. Helgason, ``Differential Geometry, Lie Groups, and
Symmetric Spaces,'' Academic Press, New York, San Diego, 1978.

\bibitem[Hi1] {Hi1}O. Hijab, Hermite functions on compact Lie groups. I,
\textit{J. Funct. Anal.} \textbf{125} (1994), 480--492.

\bibitem[Hi2] {Hi2}O. Hijab, Hermite functions on compact Lie groups. II,
\textit{J. Funct. Anal.} \textbf{133} (1995), 41--49.

\bibitem[Ho] {Ho}G. Hochschild, The structure of Lie groups. Holden-Day, San
Francisco, 1965.

\bibitem[Ki] {Ki}A. Kirillov, Geometric quantization, \textit{in} ``Dynamical
Systems IV'' (V. Arno\v{l}d and S. Novikov, Eds.). Encyclopaedia of
Mathematical Sciences, Vol. 4, Springer-Verlag, New York, Berlin, 1990.

\bibitem[KR] {KR}K. Kowalski and J. Rembieli\'{n}ski, Coherent states for a
particle on a sphere, preprint. [htttp://xxx.lanl.gov, quant-ph/9912094]

\bibitem[L1] {L1}N. Landsman, Rieffel induction as generalized quantum
Marsden-Weinstein reduction, \textit{J. Geom. Phys.} \textbf{15} (1995), 285--319.

\bibitem[L2] {L2}N. Landsman, Mathematical topics between classical and
quantum mechanics. Springer Monographs in Mathematics. Springer-Verlag, New
York, 1998.

\bibitem[LS] {LS}L. Lempert and R. Sz\"{o}ke, Global solutions of the
homogeneous complex Monge-Amp\`{e}re equation and complex structures on the
tangent bundle of Riemannian manifolds, \textit{Math. Ann.} \textbf{290}
(1991), 689--712.

\bibitem[Lo] {Lo}R. Loll, Non-perturbative solutions for lattice quantum
gravity. \textit{Nucl. Phys. B} \textbf{444} (1995), 619--639.

\bibitem[M1] {M1}J. Mitchell, Short time behavior of Hermite functions on
compact Lie groups, \textit{J. Funct. Anal.} \textbf{164} (1999), 209-248.

\bibitem[M2] {M2}J. Mitchell, Asymptotic expansions of Hermite functions on
compact Lie groups, preprint.

\bibitem[R] {R}J. Rawnsley, A nonunitary pairing of polarizations for the
Kepler problem, \textit{Trans. Amer. Math. Soc.} \textbf{250} (1979), 167--180.

\bibitem[RS] {RS}M. Reed and B. Simon, ``Methods of Modern Mathematical
Physics, I: Functional Analysis,'' Academic Press, New York, London, 1972.

\bibitem[Sa] {Sa}G. Sadasue, Equivalence-singularity dichotomy for the Wiener
measures on path groups and loop groups, \textit{J. Math. Kyoto Univ.}
\textbf{35} (1995), 653--662.

\bibitem[S1] {S1}I. Segal, Mathematical problems of relativistic physics,
Chap.\thinspace VI, \textit{in} ``Proceedings of the Summer Seminar, Boulder,
Colorado, 1960, Vol. II.'' (M. Kac, Ed.). Lectures in Applied Mathematics,
American Math. Soc., Providence, Rhode Island, 1963.

\bibitem[S2] {S2}I. Segal, Mathematical characterization of the physical
vacuum for a linear Bose-Einstein field, \textit{Illinois J. Math.} \textbf{6}
(1962), 500-523.

\bibitem[S3] {S3}I. Segal, The complex wave representation of the free Boson
field, \textit{in} ``Topics in functional analysis: Essays dedicated to M.G.
Krein on the occasion of his 70th birthday'' (I. Gohberg and M. Kac, Eds).
Advances in Mathematics Supplementary Studies, Vol. 3, pp. 321-343. Academic
Press, New York, 1978.

\bibitem[St] {St}M. Stenzel, The Segal-Bargmann transform on a symmetric
space of compact type, \textit{J. Funct. Anal.} \textbf{165} (1999), 44--58.

\bibitem[Su] {Su}H. Sugita, Holomorphic Wiener function, \textit{in} ``New
trends in stochastic analysis (Charingworth, 1994)'' (K. D. Elworthy, S.
Kusuoka, and I. Shigekawa, Eds.) pp. 399--415. World Sci. Publishing, River
Edge, NJ, 1997.

\bibitem[T] {T}T. Thiemann, Reality conditions inducing transforms for
quantum gauge field theory and quantum gravity, \textit{Classical Quantum
Gravity} \textbf{13} (1996), 1383-1403.

\bibitem[Wo] {Wo}N. Woodhouse, ``Geometric Quantization,'' Second Edition,
Oxford Univ. Press, Oxford, New York, 1991.

\bibitem[Wr] {Wr}K. Wren, Constrained quantisation and $\theta$-angles. II.
\textit{Nuclear Phys. B} \textbf{521} (1998), 471--502.

\bibitem[Z] {Z}F. Zhu, The heat kernel of the second classical domain and of
the symmetric space of a normal real form, \textit{Chinese J. Contemp. Math.}
\textbf{13} (1992), 181--200.
\end{thebibliography}
\end{document}